\documentclass[aps,pra,twocolumn,epsfig,rotate,showpacs]{revtex4}
\usepackage{amsfonts} 
\usepackage{amsmath}
\usepackage{amssymb}
\usepackage{bm}
\usepackage[usenames, dvipsnames]{color} 
\usepackage{dcolumn}
\usepackage{epic} 
\usepackage{epsfig}
\usepackage{graphicx}
\usepackage[abs]{overpic}
\usepackage[lofdepth,lotdepth,caption=false]{subfig}
\usepackage[breaklinks,colorlinks = true,linkcolor = blue,urlcolor=blue,citecolor=blue]{hyperref}
\usepackage{soul}
\usepackage{times}
\usepackage{ulem}
\usepackage{wrapfig} 
\usepackage{xy} 
\usepackage{color,soul}

\newcommand{\beq}{\begin{eqnarray}}
\newcommand{\eeq}{\end{eqnarray}}
\newcommand{\ua}{\uparrow}
\newcommand{\da}{\downarrow}
\newcommand{\dg}{\dagger}
\newcommand{\mbf}{\mathbf}
\newcommand{\la}{\langle}
\newcommand{\ra}{\rangle}
\newcommand{\bso}{\boldsymbol}
\begin{document}
\title{A Simple Hubbard Model for the Excited States of $\pi$ Conjugated -acene Molecules}
\author{Z. S. Sadeq}
\email{sadeqz@physics.utoronto.ca}
\author{J.E. Sipe}
\affiliation{Department of Physics, University of Toronto, 60 St. George Street, Toronto, Ontario, Canada, M5S 1A7}
\date{\today}
\begin{abstract}
In this paper we present a model for the electronic excited states of $\pi$ conjugated -acene molecules such as tetracene, pentacene, and hexacene. We use a simple Hubbard model with a limited basis to describe the low lying excitations with reasonable quantitative accuracy. We are able to produce semi-analytic wavefunctions for the electronic states of the system, which allows us to compute the density correlation functions for various states such as the ground state, the first two singly excited states, and the lowest lying doubly excited state. We show that in this lowest lying doubly excited state, a state speculated to play a role in the singlet fission process, the electrons and holes behave in a triplet-like manner. We also compute the two-photon absorption of these -acenes, and show that per number density it is comparable to that of other organic molecules such as coronene and hexa-peri-hexabenzocoronene (HBC).
\end{abstract}
\email{sadeqz@physics.utoronto.ca}
\maketitle

\section{Introduction}

Recent interest in the photophysics of -acene compounds \cite{zim, rch1,
rch2,rch3}, specifically tetracene ($C_{18}H_{12}$), pentacene ($C_{22}H_{14}$)
and hexacene ($C_{26}H_{16}$), is due in part to the possibility of using
these molecules to generate two triplet excitons following the absorption of
a single photon, a phenomenon known as multi-exciton generation (MEG) or
singlet fission (SF). The details of this process, potentially important for
improving solar cell efficiency \cite{michlrev}, are the subject of vigorous debate.
Some have speculated as to the role of a doubly excited state, on a single molecule, as an intermediate state in the SF process \cite{zim}. There are some reported observations of the intermediate doubly excited state in pentacene \cite{xyz1} and tetracene \cite{xyz2}, but experimental work on these systems is also
controversial. Relevant to these issues, as well as to the photophysics of
these compounds more generally, are the energies of the lowest lying excited
states, their nature, and the transition dipole matrix elements connecting them with
the ground state and with each other. Particularly important are the
lowest lying singlet and triplet states, $S_{1}$ and $T_{1}$ respectively,
and a low-lying state that involves two excited electron-hole pairs. In
this paper we consider an approach that yields simple estimates for the wave
functions of these states and the ground state, from which other properties
of interest can then be deduced.

There are a number of traditional approaches used to describe these states.
The Pariser-Parr-Pople (PPP) method employs an extended Hubbard model \cite
{ragu,mazumdar,PPP,barf,dftcalc1}. More complex methods of treating the many body wavefunction
involve configuration interactions (CI),  strategies based on density
functional theory (DFT) \cite{bend, dftcalc1,dftcalc2,zim, silb1,prezh}, as well as applications of density matrix renormalization group (DMRG) to problems in quantum chemistry \cite{hach,ragu}. In CI calculations,
one typically employs a restricted basis, considering only a small subset of
the total levels of the molecule \cite{zim}. For studies focused on the low lying electronic excitations, this can be the levels close to the highest occupied molecular
orbital (HOMO) and the lowest unoccupied molecular orbital (LUMO) of the
molecule. The methods are computationally intensive and do not produce
simple expressions for the wavefunctions. Even more complicated approaches
combine aspects of DFT and the Hubbard model \cite{dfthubb}.

The lack of simpler and more physically intuitive strategies to describe
these states has hindered the investigation of many of the optical
properties of these molecules and their lattices, the study of which would
lead to a better understanding of the states themselves. The nonlinear
optical properties of these molecules, for example, remain largely
unexplored, and with insight into the wavefunctions of the various
electronic excited states one could propose a variety of nonlinear optical
experiments, involving, for example, the coherent control of populations and
currents \cite{jr1,jr2,jr3,paul}, to probe the electronic structure further.

The approach we use in this paper borrows features from quantum chemistry strategies, but uses them in more approximate ways. It harks back to the seminal work of Ruedenberg and Platt \cite{rud,platt}, where empirical models were used to describe the ground and excited states of various conjugated molecules. Our goal is not simply to calculate the energies and wavefunctions of these states, as there are now far more sophisticated ways of doing so, but rather to provide some insight into the behavior of electrons in these states. In these $\pi $ conjugated
systems, the electrons involved in the low lying excitations of the system
come from $p_{z}$ orbitals on the carbon backbone. These electrons can hop
between each carbon atom, leading to states that are often delocalized over
the entire molecule \cite{zim}. We use a tight-binding model based on these
orbitals, with a simple Hubbard interaction as a correction to
introduce the Coulomb repulsion; this splits the energies of the singlet
and triplet states. Adopting a standard tight-binding hopping
matrix element, the model involves one adjustable parameter for each
molecule, the Hubbard interaction energy. Choosing a restricted basis, we
set this parameter to obtain agreement with the energy difference between
the $S_{1}$ state and the ground state. We find that the energy of the
other low lying states, and transition dipole moment matrix elements connecting the low
lying states, are often as well described in this simple approach as they
are by much more numerically intensive methods. The wave functions that
result are semi-analytic, and the electronic properties of the states can be
easily explored. We also introduce a simple approach to characterizing the electron-electron correlations in these states, and use it to identify the nature of the states we calculate. While the approach can be used for any conjugated organic
molecule, we use it in this first application to study the electronic
structure of tetracene, pentacene, and hexacene.

This paper is written in five sections. In Section \ref{smod} we present the model
used to describe the low lying excitations of conjugated -acenes, and in
Section \ref{sres} we outline some of the predictions for the transition energies.
In Section \ref{sden} we introduce and present electron correlation functions for
the ground state, the first singlet and triplet states, and the first doubly
excited state. In Section \ref{schi3} we calculate the two-photon absorption (TPA) of the acenes. In Section \ref{sconc} we conclude.

\section{\label{smod} Model}
We take the Hamiltonian to be given by the sum of tight-binding and Hubbard contributions
\begin{equation}
H = H_{TB} + H_{Hu},
\label{fulham}
\end{equation}
where the tight-binding Hamiltonian is
\begin{equation}
H_{TB} = -t \sum_{\langle i,j \rangle, \sigma} c^{\dagger}_{i\sigma} c_{j\sigma},
\label{TBeq}
\end{equation}
with $\sigma$ the spin label and $i,j$ site labels; the angular brackets indicates sum over nearest neighbors and the operators $c_{i\sigma}$ satisfy the usual anti-commutation relations, $\{ c_{i\sigma},c^{\dagger}_{j\sigma'}\} = \delta_{ij}\delta_{\sigma\sigma'}$. Each carbon atom is treated as a site, contributing a $p_{z}$ electron. The hopping parameter $t$, or transfer integral, is set to $t = 2.66 \text{ eV}$, the standard value for $sp^{2}$ bonded carbon \cite{cpam}.  The tight-binding Hamiltonian, Eq. (\ref{TBeq}), can be written in terms of the number operators associated with its eigenbasis of delocalized states,

\begin{equation}
H_{TB} = \sum_{m,\sigma} \hbar\omega_{m}  C^{\dagger}_{m\sigma}  C_{m\sigma},
\end{equation}
where
\begin{eqnarray}
\label{TBCdg}
&& C^{\dagger}_{m\sigma} = \sum_{i} M^{*}_{im} c^{\dagger}_{i\sigma}, \\ 
&& C_{m\sigma} = \sum_{i} M_{im} c_{i\sigma}, \label{TBC}
\end{eqnarray}
and for all $m \geq 1 $ the term $M_{im}$ is typically non-zero for all $i$. The new destruction and creation operators also obey the anti-commutation relations $\{ C_{m\sigma}, C^{\dagger}_{m'\sigma'} \} = \delta_{mm'}\delta_{\sigma\sigma'}$. The ground state of $H_{TB}$ is then
\begin{equation}
|0\rangle = \prod^{N/2}_{m=1} C^{\dagger}_{m\uparrow} \prod^{N/2}_{m'=1} C^{\dagger}_{m'\downarrow} |\text{vac}\rangle,
\label{TBgs}
\end{equation}
which we call the ``nominal vacuum," where $|\text{vac}\rangle$ represents the full vacuum and $N$ is the number of electrons in the system;  the indices $m$ are ordered so the energies $\hbar\omega_{m}$ increase with increasing $m$. The set of orbitals that are occupied in the tight-binding ground state are the filled or ``valence" orbitals and those that are not occupied in the ground state are the unfilled or ``conduction" orbitals. Each filled orbital is occupied by two electrons, constituting a ``closed shell". As usual, we denote the highest occupied orbital as the HOMO, and the $n^{th}$ state below that the HOMO-n orbital; similarly, we denote the lowest unoccupied orbital as the LUMO, and the $m^{th}$ state above that the LUMO+m orbital. 

The use of $H_{TB}$ alone as the molecular Hamiltonian would lead to a degeneracy between singlet and triplet states, which in fact is broken by the electron-electron repulsion. The simplest approach to take that into account is to include in Eq. (\ref{fulham}) a Hubbard Hamiltonian,
\begin{equation}
H_{Hu} = U \sum_{i} n_{\uparrow}(i) n_{\downarrow}(i).
\label{HubHam}
\end{equation}
Here $n_{\sigma}(i)$ is the number operator for site $i$ and spin $\sigma$, $n_{\sigma}(i) = c^{\dagger}_{i\sigma} c_{i\sigma}$ and $U > 0$ is the single Hubbard parameter introduced in this model.
\begin{figure}[h]
\begin{center}
\includegraphics[scale=0.45]{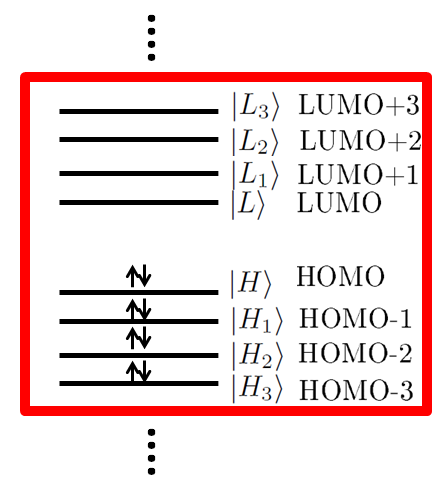}
\caption{Diagonalization of the tight-binding Hamiltonian leads to $N$ electronic levels. The tight-binding ground state (see Eq. (\ref{TBgs})) is the configuration where half the levels are filled. The red box represents the levels involved in the diagonalization of the Hamiltonian Eq. (\ref{fulham})}
\label{AS}
\end{center}
\end{figure}

We now rewrite the total Hamiltonian (\ref{fulham}) in the electron-hole basis.  Electron creation is designated by the operator $a^{\dagger}_{i \sigma}$ and hole creation by $b^{\dagger}_{i \sigma}$, with

\begin{eqnarray}
&& a_{L_{m}\sigma} \equiv C_{L_{m}\sigma},\\
&& b^{\dagger}_{H_{n}\sigma} \equiv C_{H_{n}\tilde{\sigma}},
\end{eqnarray}
where $\tilde{\sigma}$ is the opposite spin of $\sigma$. The label $L_{m}$ denotes the LUMO+m state, and the label $H_{n}$ denotes the HOMO-n state (see Fig \ref{AS}); the LUMO state itself is denoted as $L_{0}$ and the HOMO state as $H_{0}$, or for simplicity $L$ and $H$ respectively. The full form of the Hamiltonian in the electron-hole basis is given in Appendix \ref{tothubbhamsupp}. States constructed by letting one electron and one hole creation operator act on the nominal vacuum, Eq. (\ref{TBgs}), are called ``single excitations," and those constructed by letting two electron and two hole creation operators act on the nominal vacuum, Eq. (\ref{TBgs}), are called ``double excitations."


We adopt an approach from quantum chemistry and select an ``active space" defined by a set of tight-binding excited states, identified by overbars. Together with the nominal vacuum (\ref{TBgs}), these tight-binding excited states will be used to approximately diagonalize the total Hamiltonian (\ref{fulham}); the electron and hole operators that are involved in writing these tight-binding excited states are indicated in Fig. \ref{AS}.

The single excitations in our active space are of the form
\begin{equation}
|\overline{L_{m},H_{n}; \sigma}\rangle = a^{\dagger}_{L_{m}\sigma}b^{\dagger}_{H_{n}\tilde{\sigma}} |0\rangle,
\label{singbasis}
\end{equation}
where $m$ and $n$ range over $\{ 0,1,2,3 \}$. Singlet and triplet states can be constructed as superpositions of these, for example 

\begin{eqnarray}
\label{tbS1}
&& |\overline{S_{1}} \rangle = \frac{1}{\sqrt{2}} \left( a^{\dagger}_{L\uparrow}b^{\dagger}_{H\downarrow} + a^{\dagger}_{L\downarrow}b^{\dagger}_{H\uparrow} \right)|0\rangle, \\
&& |\overline{S_{4}} \rangle = \frac{1}{\sqrt{2}} \left(  a^{\dagger}_{L_{1}\uparrow}b^{\dagger}_{H_{1}\downarrow} + a^{\dagger}_{L_{1}\downarrow}b^{\dagger}_{H_{1}\uparrow}\right)|0\rangle, \label{tbS4}
\end{eqnarray}
for two of the singlet states, and
\begin{eqnarray}
\label{tbT1c}
&& |\overline{T^{a}_{1}} \rangle = \frac{1}{\sqrt{2}} \left( a^{\dagger}_{L\uparrow}b^{\dagger}_{H\downarrow} - a^{\dagger}_{L\downarrow}b^{\dagger}_{H\uparrow} \right)|0\rangle, \\
&& |\overline{T^{b}_{1}} \rangle =  a^{\dagger}_{L\uparrow} b^{\dagger}_{H\uparrow}  |0\rangle, \label{tbT1b} \\
&& |\overline{T^{c}_{1}} \rangle = a^{\dagger}_{L\downarrow} b^{\dagger}_{H\downarrow} |0\rangle,  \label{tbT1ca}\\
&& |\overline{T^{a}_{4}} \rangle = \frac{1}{\sqrt{2}} \left(  a^{\dagger}_{L_{1}\uparrow}b^{\dagger}_{H_{1}\downarrow} - a^{\dagger}_{L_{1}\downarrow}b^{\dagger}_{H_{1}\uparrow}\right)|0\rangle, \label{tbT4c}  \\
&& |\overline{T^{b}_{4}} \rangle =  a^{\dagger}_{L_{1}\uparrow} b^{\dagger}_{H_{1}\uparrow}  |0\rangle, \label{tbT4b} \\
&& |\overline{T^{c}_{4}} \rangle =  a^{\dagger}_{L_{1} \downarrow} b^{\dagger}_{H_{1} \downarrow}|0\rangle,
\label{tbT4ca}
\end{eqnarray}
for the associated triplet states \cite{agra}; here $S^{2}|\overline{S_{i}}\rangle = 0$ for $i= 1,4$, and $S^{2}|\overline{T^{j}_{i}}\rangle = 2|\overline{T^{j}_{i}}\rangle$ for $i=1,4$ and $j=a,b,c$. The double excitations are of the form
\begin{equation}
|\overline{L_{m},L_{m'};H_{n}, H_{n'}}\rangle = a^{\dagger}_{L_{m}\uparrow} a^{\dagger}_{L_{m'}\downarrow} b^{\dagger}_{H_{n}\downarrow} b^{\dagger}_{H_{n'}\uparrow}|0\rangle,
\label{dubbasis}
\end{equation}
where here $m,m',n,n'$ all range over $\{ 0,1,2,3 \}$. In the special case where $m=m'$ and $n=n'$ we write $|\overline{2L_{m}H_{n}}\rangle$ for $|\overline{L_{m}L_{m};H_{n},H_{n}}\rangle$. The lowest lying state of this type, $|\overline{2LH}\rangle$, where $m=n=0$, is sometimes referred to as $|D\rangle$ \cite{zim}. It can be written as $(a^{\dg}_{L\ua}b^{\dg}_{H\ua})(a^{\dg}_{L\da}b^{\dg}_{H\da})|0\ra$ and from (\ref{tbT1b}, \ref{tbT1ca}) can be associated with the presence of two triplets.  

When we diagonalize the Hamiltonian (\ref{fulham}) in our active space we find the Hamiltonian matrix splits into two blocks, one containing only single excitations and the second containing the nominal vacuum and the double excitations. The double excitations (\ref{dubbasis}) included in the active space are those that are coupled to the nominal vacuum by the Hubbard Hamiltonian. Other double excitations that can be constructed where $m,m',n,n'$ still range over $\{ 0,1,2,3 \}$, but where either the electrons have the same spin, or the holes have the same spin, or both, are coupled to the double excitations in our active space by the Hubbard Hamiltonian. But we find expanding the active space to include them as well leads to very small changes in our results, and so we neglect them.

Upon diagonalizing the Hamiltonian we find a new ground state that we denote by $|g\rangle$; other states are labeled, without an overbar, by the tight-binding states that contribute to them with the largest amplitude. So, for example, 
\begin{eqnarray}
|g\rangle = c^{0}_{g}|0\rangle +  \sum_{a} c^{a}_{g} |\overline{a}\rangle, \label{gsfh}
\end{eqnarray}
\beq
|S_{1} \rangle \approx c^{1}_{1} |\overline{S_{1}}\rangle + c^{4}_{1}|\overline{S_{4}}\rangle,
\eeq
\beq
|2LH\rangle =  c^{2LH}_{2LH}|\overline{2LH}\rangle + c^{0}_{2LH} |0\rangle + \sum_{a \neq 2LH} c^{a}_{2LH} |\overline{a}\rangle,
\label{states2}
\eeq
where the $c^{j}_{i}$ are complex numbers, the $|\overline{a}\rangle$ indicate double excitations (\ref{dubbasis}), and $|c^{1}_{1}| > |c^{4}_{1}|$, etc. Note that our ground state $|g\ra$ is a superposition of the nominal vacuum (a closed shell state) and double excitations (\ref{dubbasis}); the double excitations included are closed shell singlet states (where $L_{m} = L_{m'}$ and $H_{n}=H_{n'}$), as well as diradical states (where $L_{m} \neq L_{m'}$), and polyradical states (where $L_{m} \neq L_{m'}$ and $H_{n} \neq H_{n'}$). More sophisticated DMRG calculations \cite{hach} have shown that the ground states of tetracene, pentacene, and hexcacene indeed have small diradical and polyradical character; for -acenes larger than dodecacene, on the other hand, one expects a polyradical ground state. The diradical and polyradical character of the ground state can be investigated by computing the occupation numbers of the highest occupied natural orbital (HONO) and the lowest unoccupied natural orbital (LUNO) in these systems; the ``natural" orbitals are those that diagonalize the single particle density operator. For the ground states used in this paper we see the same trend observed by both Bendikov \textit{et al.}, and Hachmann \textit{et al.} \cite{hach,bend} where HONO and HONO-1 (LUNO, LUNO+1) occupation decreases (increases) as the -acene gets longer, indicating slight polyradical character. In particular, our results show good qualitative agreement with those of Hachmann \textit{et al.} For pentacene, for example, we find a HONO occupation number of 1.89, which Hachmann \textit{et al.} find 1.65 or 1.73, depending on the choice of basis set \cite{hach}. 



\begin{figure}[h]
\begin{center}
\includegraphics[scale=0.47]{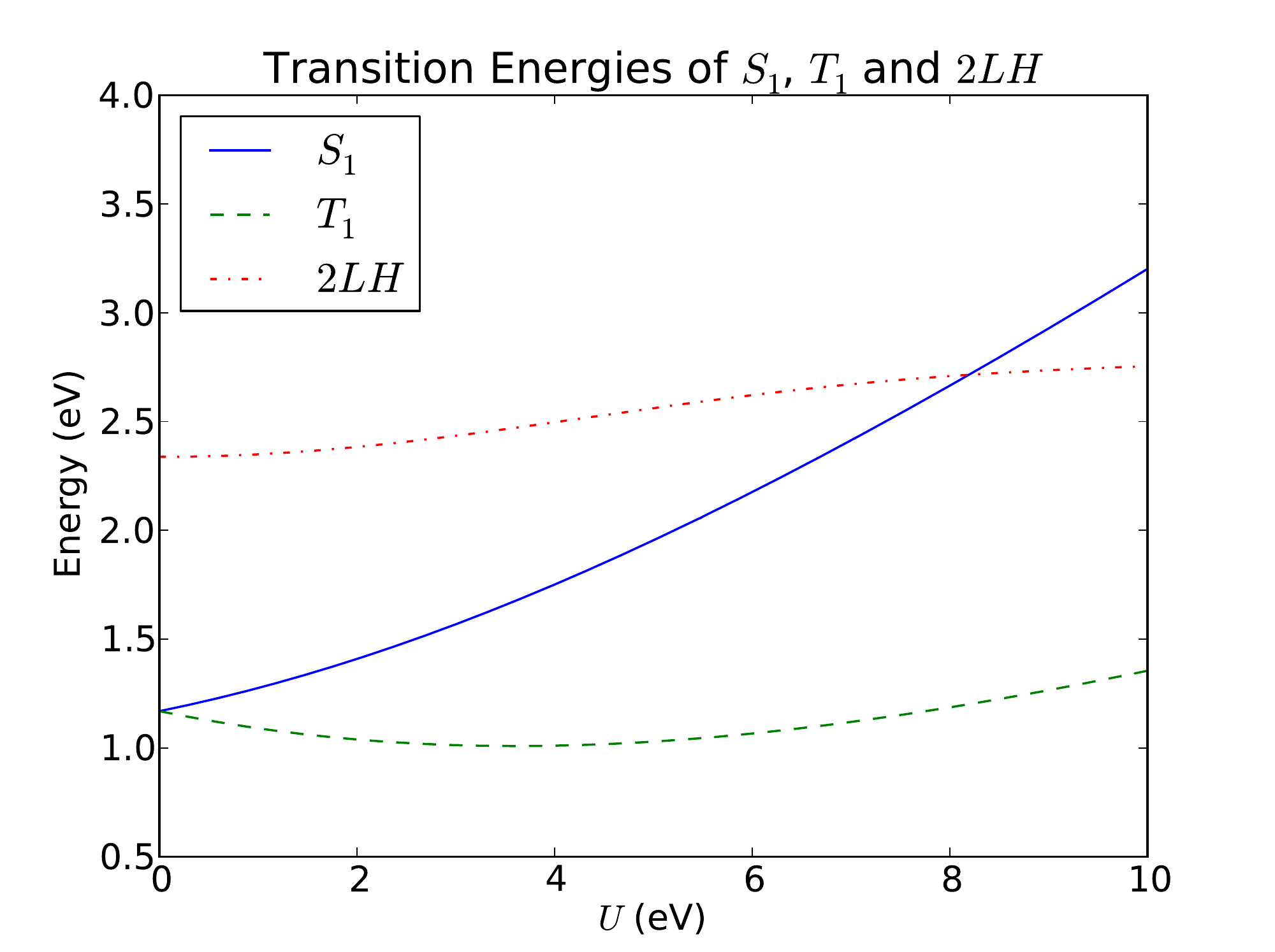}
\caption{Transition energies from the ground state to the first singlet, triplet, and doubly excited states in pentacene, as a function of the Coulomb repulsion parameter, $U$.}
\label{S1Upenplot}
\end{center}
\end{figure}

The scaling of energy levels with $U$ is shown in Fig. \ref{S1Upenplot} for pentacene; the plots for tetracene and hexacene are similar. In the tight-binding limit, $U=0$, the singlet and triplet states are degenerate; as $U$ is increased the triplet states remain degenerate but the singlet state rises faster in energy as $U$ increases, and the energy for the $2LH$ state is least sensitive to $U$. 

Taking the Hubbard energy $U$ as an adjustable parameter, a reasonable strategy would be to set $U$ so we get agreement with the experimental value of the transition energy from the ground state to $S_{1}$. Unfortunately, we are not aware of any gas phase data published for the absorption spectrum of these molecules; this would be the most appropriate for comparison with our model of isolated molecules. Available data is for molecules in solution, except for the triplet state in pentacene, where the energy is extracted from experiments on pentacene dopants in a tetracene single crystal, interpreted with the aid of an energy transfer model \cite{burg}. The early experimental data for the singlet states of tetracene and pentacene tabulated by Yamagata \textit{et al.} \cite{silb1} is for tetracene and pentacene with a solvent of benzene, and the results of Angliker \textit{et al.} \cite{hexabs} are for the singlet states of hexacene in a solvent of silicone oil. For hexacene we only report the bright singlet states, and compare to the bright singlet states measured by Angliker \textit{et al.} \cite{hexabs}.  For the triplet states in tetracene, the work was carried out by V{\"o}lcker \textit{et al.} in a solution of 2-methyltetrahydrofuran \cite{volk}, and in pentacene the first triplet state was investigated by Burgos \textit{et al.} \cite{burg}. In the absence of gas phase data, we set our values of $U$ to give the correct transition energy from the ground state to $S_{1}$ of the molecules in solution. 

\begin{widetext}
\section{\label{sres} Electronic States of Tetracene, Pentacene and Hexacene}

In Table \ref{numtabpentethex} we present the singlet energies, and the energy for $2LH$, found from our calculations; in Table \ref{oscstr} we give the oscillator strengths for the singlet transitions, where the oscillator strength is
\begin{equation}
f_{qq'} = \frac{2m_{e}\omega_{qq'}}{3\hbar e^{2}} \sum_{\beta}|\mu^{\beta}_{qq'}|^{2},
\end{equation}
with $\omega_{qq'}$ the frequency difference between states $q$ and $q'$, $\mu^{\beta}_{qq'}$  the $\beta$ component of the transition dipole matrix element between states $q$ and $q'$, and $m_{e}$ the electron mass. The form of the dipole operator used to compute the transition dipole matrix elements is given in Appendix \ref{dipsupp}. We assume the bond length to be uniform in all three molecules with a bond length $l = 1.42 \AA$ \cite{penparam}. In both tables we compare with experimental values in solution, as indicated in the figure captions, in the absence of any available gas phase data. While the energy of $S_{1}$ is in each case set to be in agreement with the experimental value by the choice of $U$, other energies such as $S_{2}$, and the oscillator strengths, can be taken as predictions of our model, which of course do not include any of the solvent corrections that could be expected in the experimental data. Especially considering this limitation, we see reasonable agreement with experimental data. 

We also find reasonable agreement with quantum chemical calculations \cite{ragu, dftcalc1,dftcalc2, silb1,prezh}, where solvent effects are also neglected; our agreement with experiment is generally comparable to that of more sophisticated quantum chemical calculations \cite{ragu, dftcalc1,dftcalc2, silb1,prezh}, where in fact different workers find very different energies for the same state. For example, for the energy of the $S_{1}$ state in pentacene, which is set in our model, the calculated values range from 2.31 eV to 3.33 eV \cite{dftcalc1,dftcalc2,silb1,prezh}. For $S_{2}$, where subject to corrections due to solvent effects our result of $4.11 \text{ eV}$ can be taken as a prediction, the results from more sophisticated calculations give values that range from $3.11 \text{ eV}$ \cite{prezh} to $4.23 \text{ eV}$ \cite{silb1}. Similarly, there are a wide range of predicted oscillator strengths from those more sophisticated calculations; for example, Yamagata \textit{et al.} \cite{silb1} calculate a value of 0.275 for the $g$ to $S_{1}$ transition in pentacene while Pedash \textit{et al.} \cite{prezh} calculate 0.184; we find 0.146. 


\begin{center}
\begin{table}[b]
\begin{tabular}{|c|c|c|c|}
\hline 
\textbf{Level} & \textbf{Tetracene (eV)} & \textbf{Pentacene (eV)} & \textbf{Hexacene (eV)}\tabularnewline
\hline 
\hline 
$S_{1}$ & 2.61(2.61) & 2.13(2.13) & 1.90(1.90)\tabularnewline
\hline 
$S_{2}$ & 4.97(4.21) & 4.11(3.58) & 3.57 (3.17) \tabularnewline
\hline 
$S_{3}$ & 4.19(4.53) & 3.91(4.07) & 3.80 (3.94) \tabularnewline
\hline 
$S_{4}$ & 5.31(5.46) & 4.49(4.34) & 3.97\tabularnewline
\hline 
$2LH$ & 3.43 & 2.61 & 2.06\tabularnewline
\hline 
\end{tabular}
\caption{Energies for various levels of tetracene, pentacene and hexacene. Energies reported for these levels are the absolute energies of these states minus the absolute energy of the ground state.  The Hubbard repulsion parameter was set to $U =5.54$ eV for tetracene, $U = 5.80$ eV for pentacene and $U = 6.38$ eV for hexacene. Experimental values of the molecules in solution are indicated in brackets; those for tetracene and pentacene are from data tabulated from the literature by Yamagata \textit{et al.} \cite{silb1}, where the solvent was benzene; the values for hexacene are found by Angliker \textit{et al.} \cite{hexabs}, where the solvent was silicone oil. We find a dark singlet state in hexacene at 2.93 eV, which Angliker \textit{et al.} find at 2.67 eV; we do not include it in this table.}
\label{numtabpentethex}
\end{table}
\end{center}

\begin{center}
\begin{table}[h]
\begin{tabular}{|c|c|c|c|c|}
\hline 
\textbf{Transition} & \textbf{Tetracene } & \textbf{Pentacene} & \textbf{Hexacene } & \textbf{Direction}\tabularnewline
\hline 
\hline 
$g\rightarrow S_{1}$ & 0.194 (0.108) & 0.146 (0.0995) & 0.104 (0.100) & $\mbf{\hat{y}}$ \tabularnewline
\hline 
$g\rightarrow S_{2}$ & 0.305 (0.0998) & 0.276 (0.0982) & 0.252 (0.100) & $\mbf{\hat{y}}$ \tabularnewline
\hline 
$g\rightarrow S_{3}$ & 2.84 (1.75) & 3.16 (2.41) & 3.39 (5.00) & $\mbf{\hat{x}}$ \tabularnewline
\hline 
$g\rightarrow S_{4}$ & 0.104 (0.155) & 0.0980 (0.243) & 0.0575 & $\mbf{\hat{y}} $\tabularnewline
\hline 
$S_{1}\rightarrow2LH$ & 0.138 & 0.0877 & 0.0307  & $\mbf{\hat{y}}$ \tabularnewline
\hline 
\end{tabular}
\caption{Table of oscillator strengths for certain transitions of tetracene, pentacene and hexacene. Experimental values of the oscillator strength of the molecules in solution are indicated by brackets; those for tetracene and pentacene are computed by Yamagata \textit{et al.} \cite{silb1} from the experimental values of the dipole matrix elements of tetracene and pentacene in a solvent of benzene; those for hexacene are measured by Angliker \textit{et al.} \cite{hexabs} in a solvent of silicone oil. The direction of the transitions are also provided; the molecular axes are shown in Fig. \ref{pendir} }
\label{oscstr}
\end{table}
\end{center}

\begin{center}
\begin{table}
\begin{tabular}{|c|c|c|c|}
\hline 
\textbf{Level} & \textbf{Tetracene (eV)} & \textbf{Pentacene (eV)} & \textbf{Hexacene (eV)}\tabularnewline
\hline 
\hline 
$T_{1}$ & 1.45 (1.35) & 1.06 (0.86) & 0.84 \tabularnewline
\hline 
$T_{2}$ & 2.67  & 2.07 & 1.69 \tabularnewline
\hline 
$T_{3}$ & 3.40 & 2.74 & 2.36 \tabularnewline
\hline 
$T_{4}$ & 3.70 & 3.02 & 2.52 \tabularnewline
\hline 
\end{tabular}
\caption{Energies of triplet levels in tetracene, pentacene and hexacene. Energies reported for these levels are the absolute energies of these states minus the absolute energy of the ground state. The experimental value for the first triplet state of tetracene is taken from V{\"o}lcker \textit{et al.} \cite{volk}, where tetracene was studied in a solvent of 2-methyltetrahydrofuran, while that of pentacene is taken from Burgos \textit{et al.} \cite{burg} where pentacene molecules were studied as dopants in tetracene single crystals.}
\label{tripabspentethex}
\end{table}
\end{center}


\begin{figure}[h]
\begin{center}
\includegraphics[scale=0.315]{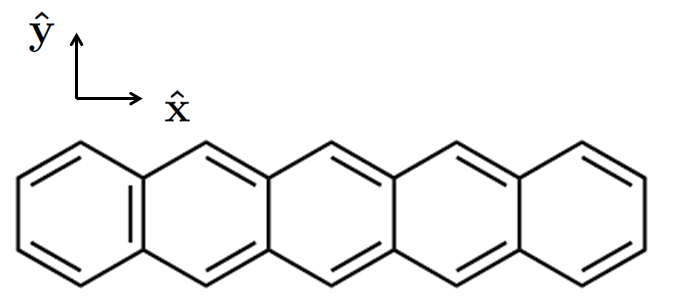}
\caption{A cartoon representation of pentacene showing the molecular axes; the long axis is denoted as the $\mbf{\hat{x}}$ axis, while the short axis is denoted as the $\mbf{\hat{y}}$ axis. }
\label{pendir}
\end{center}
\end{figure}

\begin{figure}[b]
\begin{center}
\includegraphics[scale=0.35]{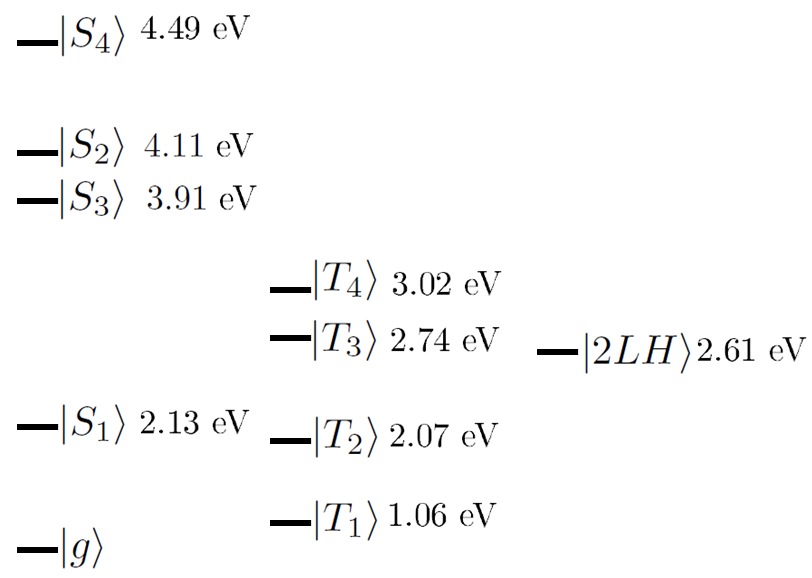}
\caption{Level structure of pentacene showing the first four singlet and triplet excited states as well as the first doubly excited state.}
\label{penlvlsfig}
\end{center}
\end{figure}

The lowest energy triplets predicted by this model are presented in Table \ref{tripabspentethex}; the triplet energies predicted by our model show good quantitative agreement with those of a more sophisticated calculation \cite{hach}. The level structure for pentacene is shown in Fig. \ref{penlvlsfig}.  Note that we follow Yamagata \textit{et al.} \cite{silb1} and assign levels by comparing our calculated oscillator strengths with those determined experimentally, rather than by comparing our calculated energies with those determined experimentally; this leads to $S_{3}$ having a slightly lower energy than $S_{2}$ in tetracene and pentacene. Our convention for labeling the $S_{2}$ and $S_{3}$ states also ensures these states have similar tight-binding excitation contributions over the three molecules that we study.

The singlet and triplet states are composed of states in which there is a hole in $H_{i}$ and an electron in $L_{j}$; we denote this by $H_{i} \rightarrow L_{j}$, identifying the motion of an actual electron necessary to create the single excitation from the nominal vacuum. In all molecules $S_{1}$ is made up of mostly $H \rightarrow L$. The largest contribution to $S_{2}$ is from $H_{1} \rightarrow L_{1}$ in all molecules. However, in tetracene there is significant mixing with $H_{3} \rightarrow L$ and $H \rightarrow L_{3}$, in pentacene there is significant mixing with $H_{2} \rightarrow L$, $H \rightarrow L_{2}$, $H_{3} \rightarrow L$ and $H \rightarrow L_{3}$, and in hexacene there is significant mixing with $H_{2} \rightarrow L$ and $H \rightarrow L_{2}$. The state $S_{3}$ is composed only of $H_{2} \rightarrow L$ and $H \rightarrow L_{2}$ in tetracene, only of $H_{2},H_{3} \rightarrow L$ and $H \rightarrow L_{2},L_{3}$ in pentacene, and only of $H_{3} \rightarrow L$ and $H \rightarrow L_{3}$ in hexacene. In all molecules, the largest contribution to $S_{4}$ is from $H_{1} \rightarrow L_{1}$, but in tetracene there is significant mixing with $H_{3} \rightarrow L$ and $H \rightarrow L_{3}$, in pentacene there is significant mixing with $H_{2}H_{3} \rightarrow L$ and $H \rightarrow L_{2}, L_{3}$, and in hexacene there is significant mixing with $H_{2} \rightarrow L$ and $H \rightarrow L_{2}$. 

Turning to the triplets, in all molecules the state $T_{1}$ is mainly $H \rightarrow L$ and the states $T_{2}$, $T_{3}$ are made up of mostly $H_{1} \rightarrow L$ and $H \rightarrow L_{1}$. While $T_{4}$ consists mainly of $H_{1} \rightarrow L_{1}$ in all three molecules, there is significant mixing with $H \rightarrow L_{3}$ and $H_{3} \rightarrow L$ in tetracene, with $H \rightarrow L_{2}, L_{3}$ and $H_{2},H_{3} \rightarrow L$ in pentacene, and with $H \rightarrow L_{2}$ and $H_{2} \rightarrow L$ in hexacene.

\end{widetext}
\section{\label{sden} Electron Density Correlation Function}
\subsection{Ground State Electron Correlation}
We now turn to the characterization of the states, and the identification of the impact of the Hubbard Hamiltonian on their nature. From a perspective of condensed matter physics, the most natural way to begin this is to consider the electron correlations. There one typically introduces a density operator $n(\mbf{r}) = \sum_{\sigma} n_{\sigma}(\mbf{r})$ where $n_{\sigma}(\mbf{r})$ is the density operator associated with spin $\sigma = \ua, \da$ and $n_{\sigma}(\mbf{r}) = \sum_{s} \psi^{\dg}(\mbf{r},s) \psi(\mbf{r},s)$ where the field operator $\psi(\mbf{r},s) = \sum_{\alpha,\sigma} \chi_{\sigma}(s) \phi_{\alpha}(\mbf{r})c_{\alpha\sigma}$, with $\chi_{\sigma}(s)$ a complete set of spinor functions, $\phi_{\alpha}(\mbf{r})$ a complete set of wave functions, and the $c_{\alpha\sigma}$ fermion operators satisfying $\{c_{\alpha\sigma}, c_{\alpha'\sigma'} \} = 0$, $\{c_{\alpha\sigma},c^{\dg}_{\alpha'\sigma'} \} = \delta_{\alpha,\alpha'}\delta_{\sigma\sigma'}$. In the usual way \cite{pines,qliq} we find
\beq
n_{\sigma}(\mbf{r})n_{\sigma'}(\mbf{r'}) = \delta_{\sigma\sigma'}\delta(\mbf{r}-\mbf{r'})n_{\sigma}(\mbf{r}) + F_{\sigma\sigma'}(\mbf{r},\mbf{r'}),
\eeq
where the right-hand side follows from normal ordering the operators on the left-hand side and used the assumed completeness of the functions $\phi_{\alpha}(\mbf{r})$ and that of the spinors $\chi_{\sigma}(s)$; here
\begin{widetext}
\beq
F_{\sigma\sigma'}(\mbf{r},\mbf{r'}) = \sum_{\alpha_{1},\alpha_{2},\alpha_{3},\alpha_{4}} \phi^{*}_{\alpha_{1}}(\mbf{r})\phi^{*}_{\alpha_{3}}(\mbf{r'})\phi_{\alpha_{2}}(\mbf{r})\phi_{\alpha_{4}}(\mbf{r'})c^{\dg}_{\alpha_{3}\sigma'}c^{\dg}_{\alpha_{1}\sigma}c_{\alpha_{2}\sigma} c_{\alpha_{4}\sigma'}.
\eeq
The expectation value of the density-density correlation function in any pure or mixed state is then given by 
\beq
\la n(\mbf{r}) n(\mbf{r'}) \ra = \delta(\mbf{r}-\mbf{r'})\la n(\mbf{r})\ra + \la n(\mbf{r}) \ra \la n(\mbf{r'}) \ra g^{(2)}(\mbf{r},\mbf{r'}),
\label{dendencorr}
\eeq
\end{widetext}
where the correlation function
\beq
g^{(2)}(\mbf{r},\mbf{r'}) = \frac{\sum_{\sigma,\sigma'} \la F_{\sigma\sigma'}(\mbf{r},\mbf{r'}) \ra}{\la n(\mbf{r}) \ra \la n(\mbf{r'}) \ra}.
\label{g1ig}
\eeq
The first term on the right hand side of Eq. (\ref{dendencorr}) is associated with the contribution of the ``same" electron in $n(\mbf{r})$ and $n(\mbf{r'})$; the second term, characterized by the dimensionless quantity $g^{(2)}(\mbf{r},\mbf{r'})$, describes the correlation of pairs of electrons above and beyond what one would expect simply from the varying densities. For some models, such as a noninteracting, or ideal, Fermi gas at zero temperature, it can be easily evaluated \cite{qliq}.

\begin{figure}[h]
\begin{center}
\includegraphics[scale=0.4]{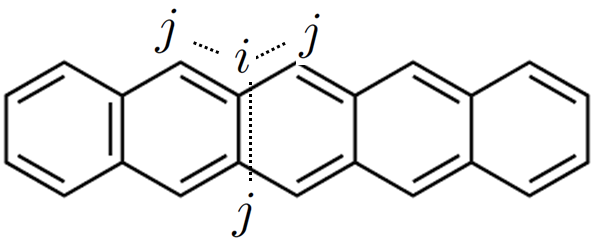}
\includegraphics[scale=0.4]{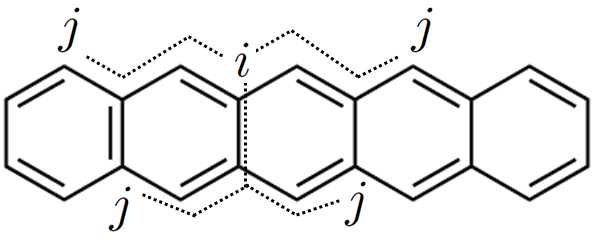}
\caption{A cartoon figure of pentacene and the separation of sites by different bond length distances. In the top figure, we indicate the procedure to compute $g^{(2)}(s = 1)$. For the particular site $i$ there are three sites $j$ that contribute for $s = 1$; we sum these contributions, then sum over all sites $i$, and finally normalize appropriately. The bottom figure shows the procedure for $s = 3$. For site $i$ the contributions from the four sites that contribute to $g^{(2)}(s = 3)$ are shown in the figure.} 

\label{g1expln}
\end{center}
\end{figure}

Here we want to associate a density operator $n_{\sigma}(i)$ for an electron with spin $\sigma$ at a particular site $i$,

\begin{eqnarray}
&& n_{\sigma}(i) = c^{\dg}_{i\sigma}c_{i\sigma} = \sum_{mm'}\Gamma_{mm'} (i) C^{\dagger}_{m\sigma}C_{m'\sigma},
\end{eqnarray}
where from the inverse of Eqs. (\ref{TBCdg},\ref{TBC}) we have
\begin{equation}
\Gamma_{mm'} (i) = M^{*}_{im} M_{im'},
\label{gammnm}
\end{equation}
and a total density operator at site $i$
\beq
n(i) = \sum_{\sigma} n_{\sigma}(i).
\eeq
Following the strategy for the more general discussion in the previous paragraph, we find 
\beq
n_{\sigma}(i) n_{\sigma'}(j) = \delta_{\sigma\sigma'}\delta_{ij}n_{\sigma}(i) + F_{\sigma\sigma'}(i,j),
\eeq
where we have used the fact that $M$ is a unitary matrix, and now
\beq
F_{\sigma\sigma'}(i,j) = \sum_{mm'pp'} \Gamma_{mm'}(i) \Gamma_{pp'}(j) C^{\dg}_{p\sigma'} C^{\dg}_{m\sigma} C_{m'\sigma} C_{p'\sigma'}.
\eeq
In any pure or mixed state we then have
\beq
\la n(i)n(j) \ra = \delta_{ij} \la n(i) \ra + \la n(i) \ra \la n(j) \ra g^{(2)}(i,j),
\eeq
where now the dimensionless quantity
\begin{equation}
g^{(2)}(i,j) = \frac{\sum_{\sigma\sigma'} \la F_{\sigma\sigma'}(i,j) \ra}{\langle n(i) \rangle \langle n(j) \rangle},
\label{g1def}
\end{equation}
characterizes the correlations between pairs of electrons. As in the more general discussion in the previous paragraph, in simple models this can be worked out explicitly. For example, if we neglect the Hubbard part of the Hamiltonian, Eq. (\ref{fulham}), and consider the tight-binding ground state, we have
\begin{equation}
g_{TB}^{(2)}(i,j) = 1 - \frac{\sum_{lk}\Gamma_{lk}(i) \Gamma_{kl} (j)}{2\sum_{l} \Gamma_{ll} (i) \sum_{k} \Gamma_{kk} (j)},
\label{g1tb}
\end{equation}
where the sums over $k$ and $l$ extend over all filled states. 

Although tabulating $g^{(2)}(i,j)$ for all pairs of sites $i$ and $j$ could be done, this would provide a surfeit of information. Recall that for a uniform electron gas $g^{(2)}(\mbf{r},\mbf{r'})$ only depends on $|\mbf{r}-\mbf{r'}|$, and the correlation function can be identified by giving its dependence only on that one variable. Here things are more complicated, for $g^{(2)}(i,j)$ does not depend just on the distance between $i$ and $j$, but also on where the sites $i$ and $j$ are located on the molecule. Further, since in our model electrons can move from one site $i$ to another $j$ by moving from one carbon site to another, arguably the physically relevant distance between $i$ and $j$ is not the actual distance between the sites but rather the minimum number of bond length steps necessary to get from $i$ to $j$, $s_{ij}$, which we call the \textit{bond length distance} between the sites; clearly $s_{ji} = s_{ij}$ (see Fig. \ref{g1expln}). While $g^{(2)}(i,j)$ does not depend just on $s_{ij}$, we can construct an average $g^{(2)}(s)$ for the molecule by averaging over all pairs $(i,j)$ of sites in the molecule with the same bond length distance between them. More precisely, we take
%
\begin{equation}
g^{(2)}(s) \equiv \frac{1}{\mathcal{N}(s)} \sum_{\substack{(i,j) \text{ such that } \\ s_{ij} = s}} g^{(2)}(i,j),
\label{g1fs}
\end{equation}
where $\mathcal{N}(s)$ is the number of pairs $(i,j)$ of sites appearing in the sum. Here $s= \{0,1,2, \ldots, s_{\text{max}} \}$, with $s_{\text{max}}$ is the maximum bond length distance between pairs of sites in the molecule; $s_{\text{max}}$ is dependent on the molecule, and for pentacene, for example, we have $s_{\text{max}} = 11$. 

\begin{figure}[h]
\begin{center}
\includegraphics[scale=0.45]{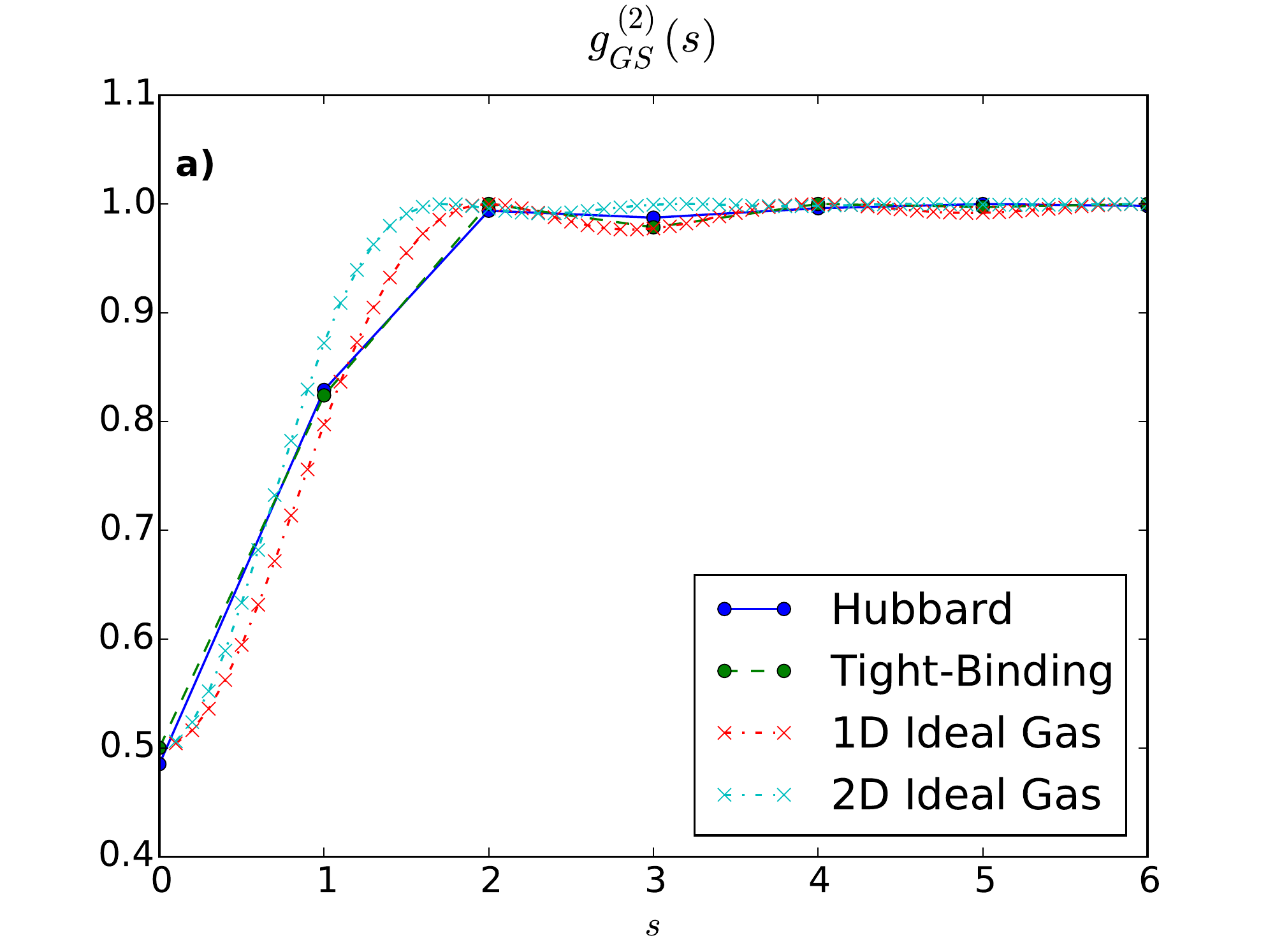}
\includegraphics[scale=0.45]{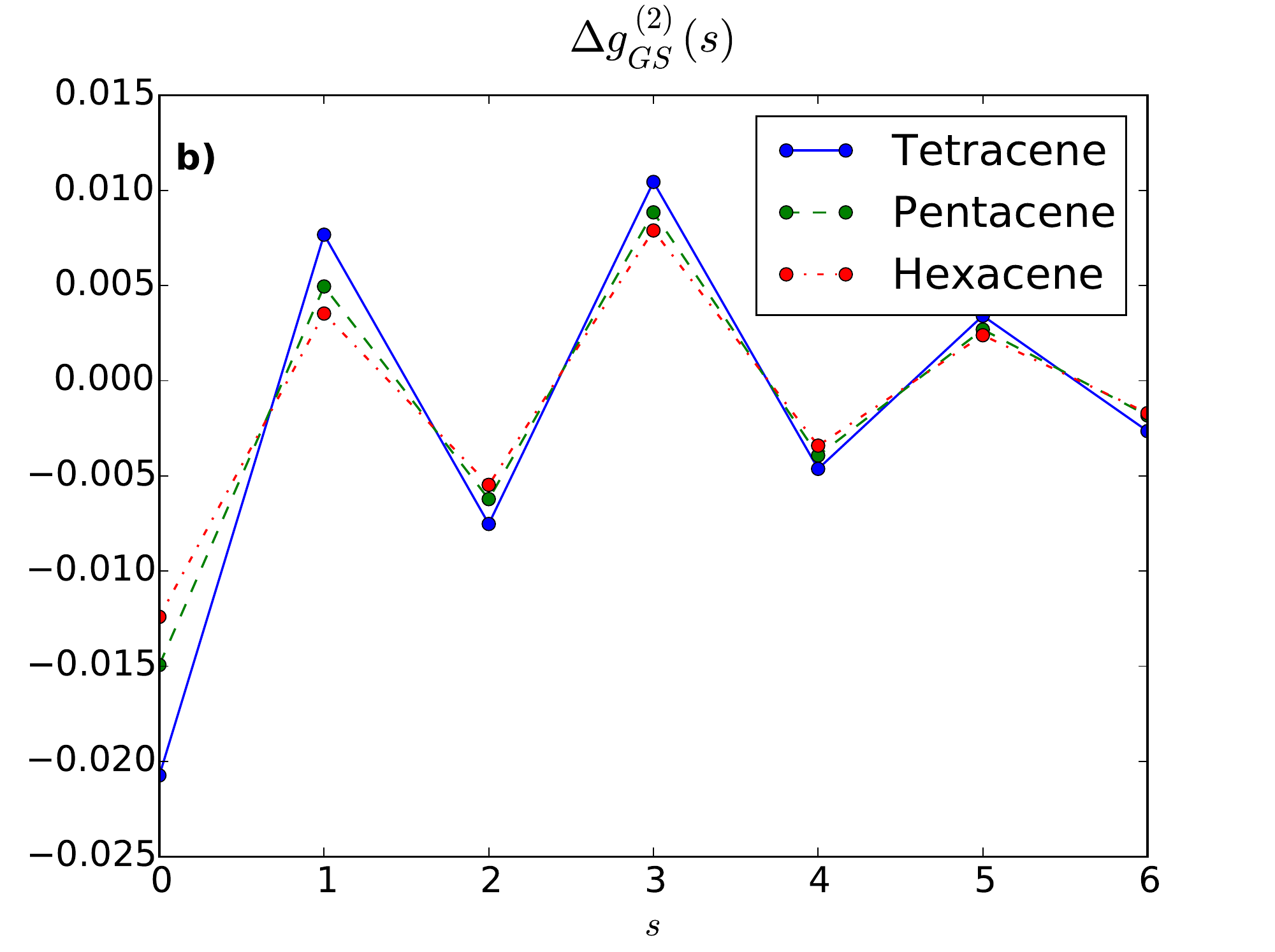}
\caption{a) The function $g_{GS}^{(2)}(s)$  is plotted for the ground state in both tight-binding and Hubbard models for pentacene, along with the corresponding correlation functions for 1D and 2D ideal gases. The Hubbard and tight-binding results are essentially indistinguishable on this scale. The quantities for tetracene and hexacene are similar. b) The difference between the two cases is illustrated in the quantity $\Delta g^{(2)}_{GS}(s) = g_{Hu;GS}^{(2)}(s) - g_{TB;GS}^{(2)}(s)$  where all three molecules are shown. Both these are plotted as a function of $s$, which identifies the distance in multiples of the bond length.}
\label{g1GS}
\end{center}
\end{figure}

To calculate $g^{(2)}(s)$ for a particular $s$, we begin with a site $i$ and find all sites $j$ with $s_{ij} = s$; we sum $g^{(2)}(i,j)$ over those. We do this for all $i$ and add the contributions. In doing so we have counted each pair $(i,j)$ twice, but we then divide by the number of contributions and recover (\ref{g1fs}). 
In Fig. \ref{g1GS} we plot $g^{(2)}(s)$ calculated from the tight-binding ground state, which we label $g^{(2)}_{TB;GS}(s)$, where the Hubbard Hamiltonian (\ref{HubHam}) is neglected, and $g^{(2)}(s)$ calculated for the ground state with the inclusion of the Hubbard Hamiltonian, which we label $g^{(2)}_{Hu;GS}(s)$. We also define
\begin{equation}
\Delta g^{(2)}_{GS}(s) = g^{(2)}_{Hu;GS}(s) - g^{(2)}_{TB;GS}(s).
\label{Deltag1}
\end{equation}
This difference is also plotted in Fig. \ref{g1GS}.

Even at the tight-binding level there is a Fermi hole at $s = 0$, analogous to the behavior of $g^{(2)}(\mbf{r},\mbf{r'})$ of Eq. (\ref{g1ig}) in an ideal Fermi gas, with $g^{(2)}(0) = 0.5$. The oscillatory behavior as a function of $s$ in $g^{(2)}_{TB;GS}(s)$ is analogous to the oscillation seen in $g^{(2)}(\mbf{r},\mbf{r'})$ as a function of $|\mbf{r}-\mbf{r'}|$ for an ideal Fermi gas \cite{pines,qliq}. The density correlation function for an ideal Fermi gas in one dimension is given by \cite{qliq}
\beq
g^{(2)}_{1D}(r) = 1 - \frac{1}{4k^{2}_{F}r^{2}} + \frac{\cos(2k_{F}r)}{4k^{2}_{F}r^{2}},
\label{1dgas}
\eeq
where we have defined $r \equiv |\mbf{r}-\mbf{r'}|$ and $k_{F}$ is the Fermi wavevector. In one dimension, the wavevector $k_{F}$ can be expressed as $k_{F} = \pi n /2$ where $n$ is the linear density. In our model this density is one electron per unit length, $n = 1 /l$, where $l$ is the bond distance. Therefore, we would identify an effective $k_{F} = \pi/2l$. One can also physically think of $k_{F}$ as being related to the wavelength of oscillation of the HOMO state. The HOMO tight-binding wavefunction has a wavelength of approximately $4l$. That is, if one is constrained to move within the sites then regions of positive and negative amplitude are separated by $4l$. This also leads to the identification of an effective $k_{F} = \pi/2l$. So either approach leads us to expect oscillations with a period of $2l$, which is indeed observed in the plots of $g^{(2)}(s)$ in Fig. \ref{g1GS}, where we also plot $g^{(2)}_{1D}(sl)$ and $g^{(2)}_{2D}(sl)$ as a function of $s$ where $g^{(2)}_{2D}(sl)$ is the density correlation function for a two dimensional ideal Fermi gas. For $g^{(2)}_{1D}(sl)$ (Eq. (\ref{1dgas})) we take the density $n = 1/l$, while for $g^{(2)}_{2D}(sl)$ we take the density to be the areal density of $p_{z}$ electrons in a graphene lattice with the same $l$. 

It is evident from Fig. \ref{g1GS} that including the Hubbard interaction reduces the electron density correlation function at $s = 0$, deepening the Fermi hole. We note that even if the size of the basis set is decreased, the behavior of $\Delta g^{(2)}_{GS}(s)$ remains qualitatively the same. Thus, even a smaller basis can effectively capture the physics of the ground state of the system. However, the larger basis we use here leads to better agreement with experiment of both the energies and the oscillator strengths. We expand on this in Appendix \ref{activespace}. 

Of the three -acenes, the tight-binding prediction of the HOMO-LUMO gap is closest to the singlet transition energy of hexacene, and furthest from that of tetracene. In Fig \ref{g1GS}, we see that it is tetracene that is most affected by the Hubbard corrections, while hexacene is the least affected, as evidenced by the magnitude of the corrections in $\Delta g^{(2)}_{GS}(s)$ for small $s$. 


\subsection{Electron-Hole Correlations in Single and Double Excited States}
\subsubsection{Electron Picture}
We now calculate the electron density correlation (\ref{g1def}) for three selected excited states: the first singlet state $S_{1}$, the first triplet state $T_{1}$, and the doubly excited state $2LH$.  

We begin with these states in the their tight-binding limit, $\overline{S}_{1}$, $\overline{T}_{1}$, and $\overline{2LH}$. To characterize the difference of the electron correlation in these states from that in the tight-binding ground state, we define 
\beq
\delta g^{(2)}_{TB;X} (s) = g^{(2)}_{TB;X}(s) - g^{(2)}_{TB;GS}(s),
\label{dg2tb}
\eeq
where $g^{(2)}_{TB;X}(s)$ is the quantity (\ref{g1fs}) computed for a particular tight-binding state ($X = \overline{S}_{1}, \overline{T}_{1},\overline{2LH}$) and $g^{(2)}_{TB;GS}(s)$ refers to (\ref{g1fs}) computed for the tight-binding ground state (\ref{TBgs}). 



\begin{figure}[h]
\begin{center}
\includegraphics[scale=0.45]{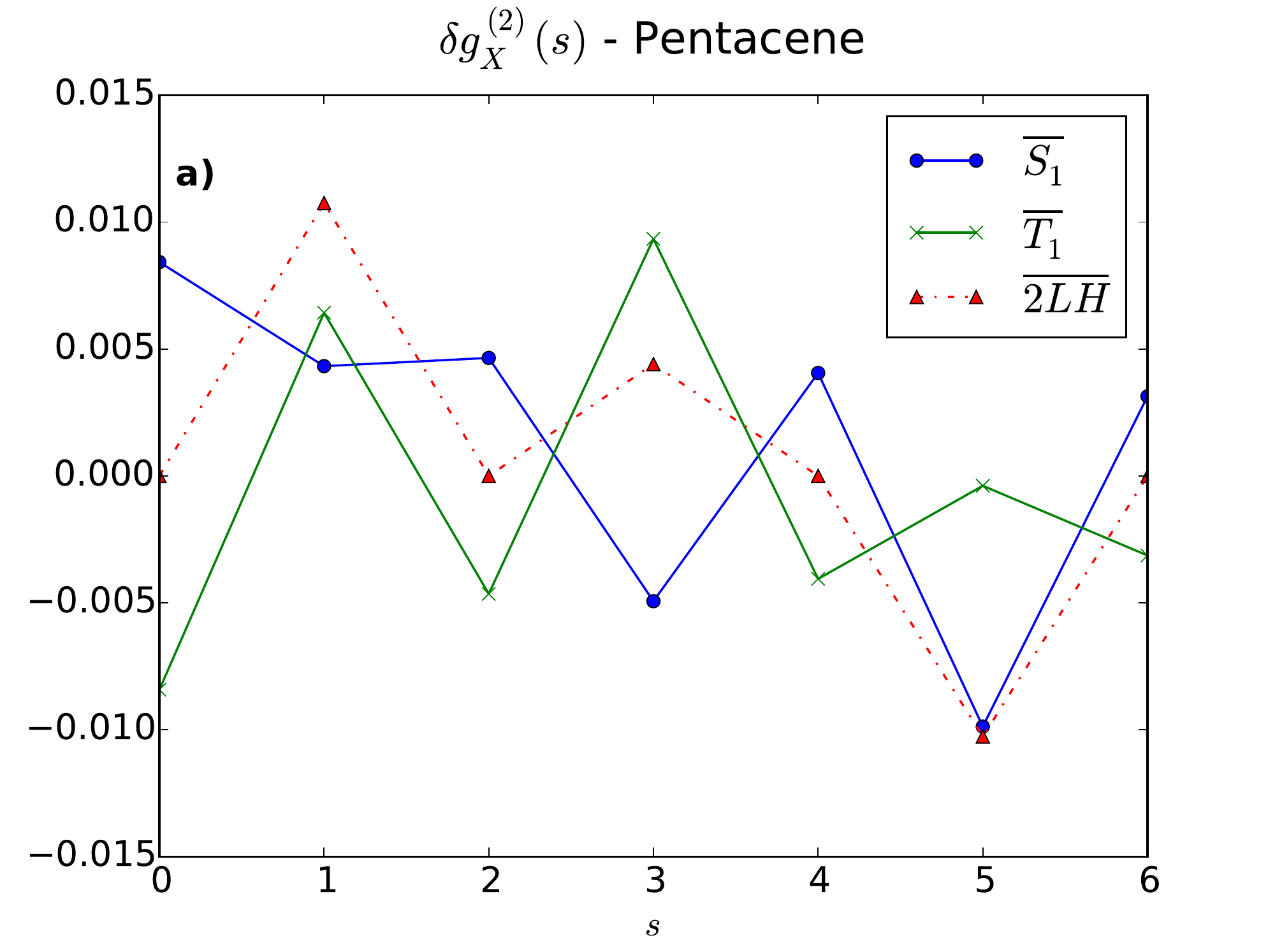}
\includegraphics[scale=0.45]{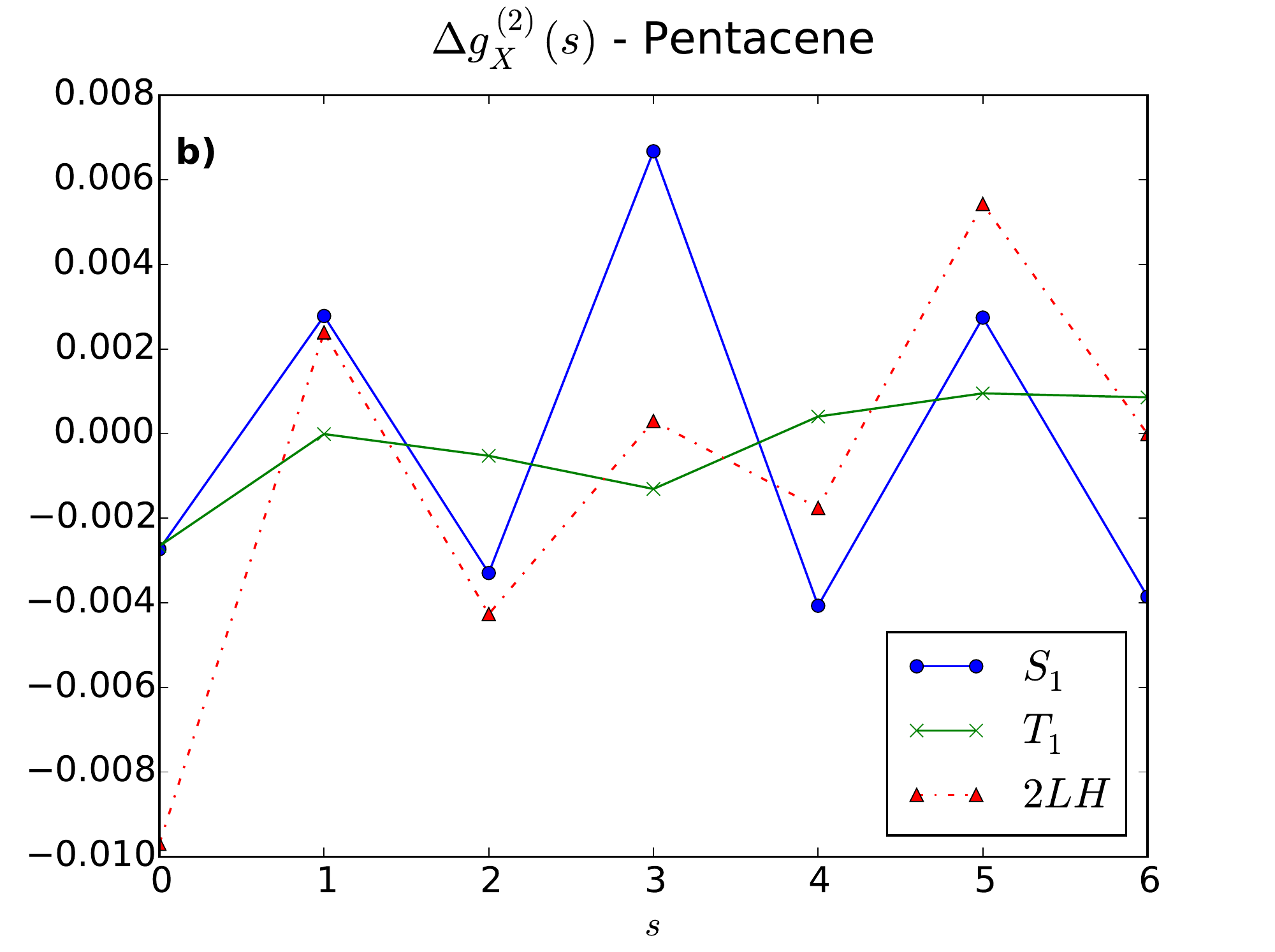}
\caption{a) The measure $\delta g^{(2)}_{X}$ plotted for $X = \overline{S}_{1}, \overline{T}_{1}, \overline{2LH}$ in pentacene.  b) The measure $\Delta g^{(2)}_{X}(s)$ plotted for the first singlet state $S_{1}$, the first triplet state $T_{1}$, and the first doubly excited $2LH$ in pentacene.}
\label{pmfig1}
\end{center}
\end{figure}

Singlet (triplet) states typically have a spatial component of their wavefunction symmetric (antisymmetric) with respect to exchange of particle coordinates, leading to a larger (smaller) spatial overlap of electrons in the singlet (triplet) states, and this is true in our model. In Fig. \ref{pmfig1}a) we plot $\delta g_{X}^{(2)}(s)$ for our states of interest in pentacene, where it is is clear that the tight-binding singlet (triplet) state has a shallower (deeper) Fermi hole relative to the tight-binding ground state. In the presence of the Hubbard interaction this splits the energy degeneracy of the tight-binding singlet and triplet states, and since that interaction also modifies the electron motion we find that the Fermi holes of $S_{1}$ and $T_{1}$ are deeper than those of $\overline{S}_{1}$ and $\overline{T}_{1}$, respectively, as is shown in Fig. \ref{pmfig1}b) where we plot $\Delta g_{X}^{(2)}(s)$ (\ref{Deltag1}) for the $X = S_{1}$, $T_{1}$ and $2LH$ states of pentacene. This difference quantity is plotted for all three -acenes for the $S_{1}$ and $T_{1}$ states in Appendix \ref{physmean}. As for the ground state, the magnitude of the Hubbard corrections $\Delta g^{(2)}_{X}(s)$ are largest for tetracene and smallest for hexacene. 

The $2LH$ state behaves similarly to the ground state. The quantities $\delta g^{(2)}_{2LH}(s)$ (\ref{dg2tb}) and $\Delta g_{2LH}^{(2)}(s)$ (\ref{Deltag1}) in pentacene are plotted in  Fig. \ref{pmfig1}a) and \ref{pmfig1}b) respectively. Note that $\delta g^{(2)}_{2LH}(0)$ is zero, indicating that the Fermi hole is no deeper or shallower in the $\overline{2LH}$ state than in the tight-binding ground state; this is due to the symmetry properties of the tight-binding eigenstates. The quantity $\Delta g^{(2)}_{2LH}(0)$ is negative, as it is for the ground state; the introduction of the Coulomb repulsion deepens the Fermi hole. The difference quantity, $\Delta g^{(2)}_{2LH}(s)$ is plotted for all three -acenes in Appendix \ref{physmean}. As for the ground state and for the singlet and triplet state, the magnitude of the Hubbard corrections $\Delta g^{(2)}_{2LH}(s)$ are largest for tetracene and smallest for hexacene. 

For the $2LH$ state, reducing the size of the basis leads to qualitatively similar behavior for the $\Delta g^{(2)}_{2LH}(s)$; the energy difference between $2LH$ and the ground state also remains similar. Thus the essential physics of this state can be effectively captured by a smaller basis. We expand on this in Appendix \ref{activespace}. 

\subsubsection{Electron Hole Picture}
\begin{figure}[h]
\begin{center}
\includegraphics[scale=0.45]{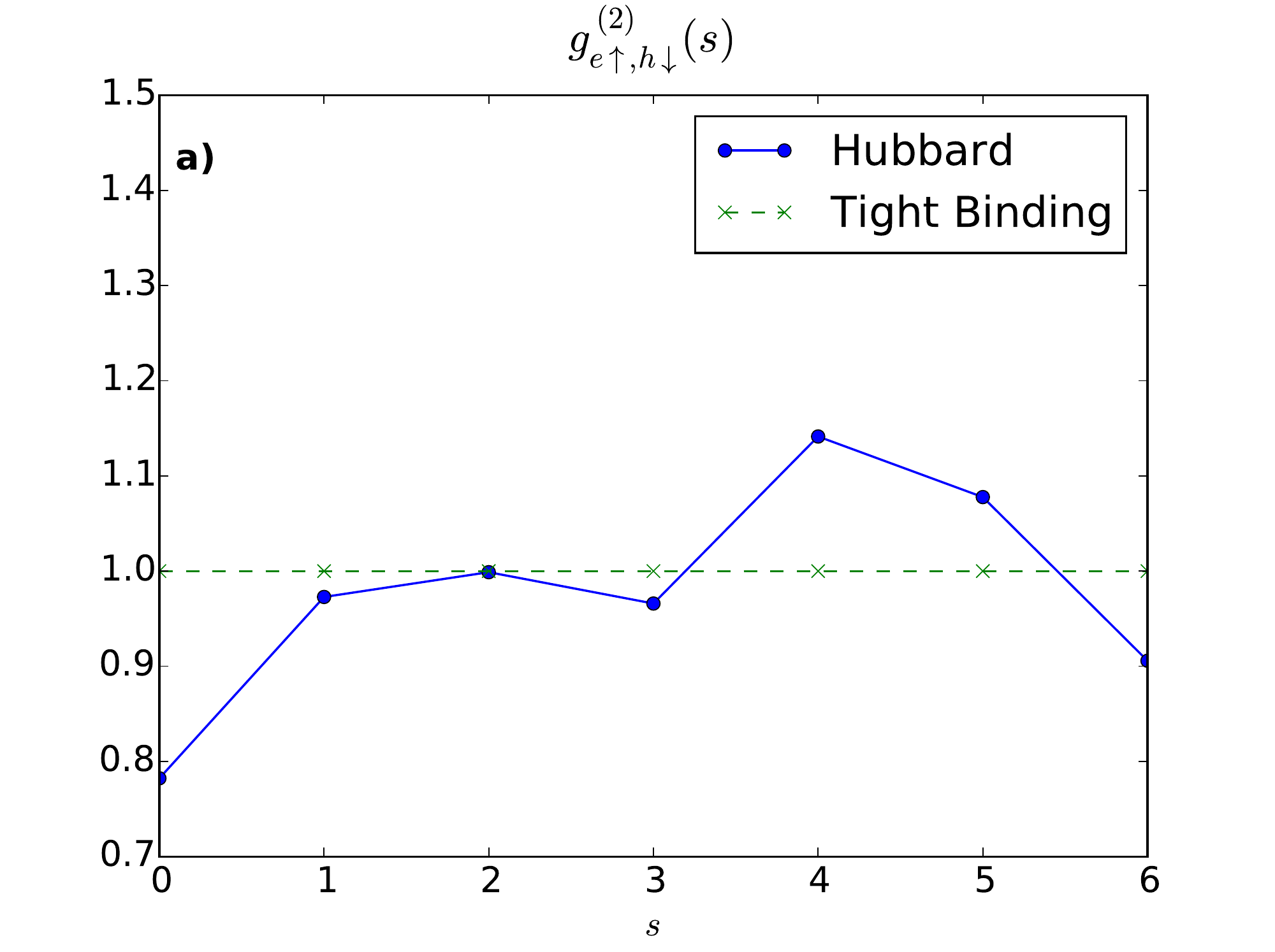}
\includegraphics[scale=0.45]{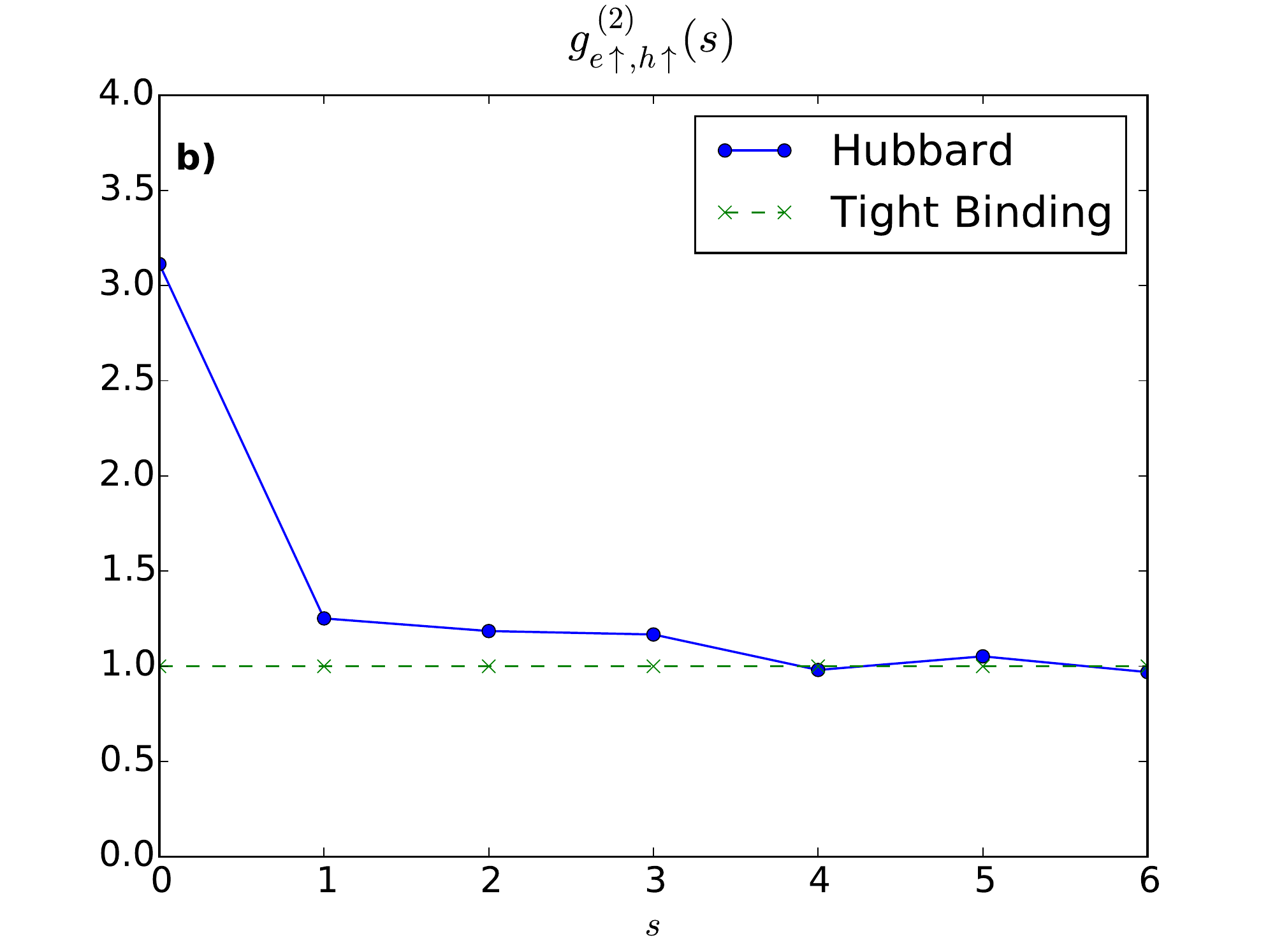}
\caption{a) $g^{(2)}_{e\uparrow h\downarrow}(s)$ and b) $g^{(2)}_{e\uparrow h\uparrow}(s)$ plotted as a function of electron-hole separation $s$ for the first doubly excited state $2LH$, in pentacene. The steps are all scaled by the bond length $l$.}
\label{ehcorrfig}
\end{center}
\end{figure}

The correlation function $g^{(2)}(s)$ involves all the electrons in the molecule, including many that in the tight-binding limit remain in the same single-particle states they inhabit in the ground state. So to identify the behavior of charges in the excited states we move to an electron-hole picture, which naturally focuses on the excitations. Of course, even the ground state $|g\ra$, calculated with the inclusion of the Hubbard Hamiltonian, includes the virtual excitation of electron-hole pairs. However, we find that their populations in the ground state are very small, and hence we can expect that the correlations of the electron and hole densities in the excited states do reliably characterize the nature of those states. The electron and hole densities are
\begin{eqnarray}
\label{enum}
n_{e\sigma}(i) = \sum_{mm'} \Gamma_{mm'}(i) a^{\dagger}_{m\sigma} a_{m'\sigma},
\end{eqnarray}
and
\begin{eqnarray} 
n_{h\sigma}(i) = \sum_{mm'} \Gamma_{mm'}(i) b^{\dagger}_{m\sigma} b_{m'\sigma},
\label{hnum}
\end{eqnarray}
and so the functions that track correlations between electrons and holes of different spins are
\begin{eqnarray}
&& g^{(2)}_{e\uparrow h\downarrow}(i,j) = \frac{\langle n_{e\uparrow}(i) n_{h\downarrow} (j) \rangle}{\langle  n_{e\uparrow}(i) \rangle \langle  n_{h\downarrow} (j) \rangle}, \label{g1ehform}
\end{eqnarray}
while those that track correlations between electrons and holes of the same spin are
\begin{eqnarray}
&&  g^{(2)}_{e\uparrow h\uparrow}(i,j) = \frac{\langle n_{e\uparrow}(i) n_{h\uparrow} (j) \rangle}{\langle  n_{e\uparrow}(i) \rangle \langle  n_{h\uparrow} (j) \rangle}. \label{g1ehssform}
\end{eqnarray}
We form the average quantities $g^{(2)}_{e\ua h\da}(s)$ and $g^{(2)}_{e\ua h\ua}(s)$ as a function of the number of bond lengths $s$, as outlined in the previous section, and evaluate these quantities in both the tight-binding limit and with the inclusion of the Hubbard Hamiltonian. We focus on the doubly excited $2LH$ state, and show the results in Fig. \ref{ehcorrfig}. 



There has been some speculation \cite{zim,xyz1} that the doubly excited state is somehow related to the singlet fission process in a dimer. It is conjectured that upon excitation to the $S_{1}$ state, the excitation relaxes down to the $2LH$ state, which breaks up into two triplet excitons, one in each molecule in the dimer. In the tight-binding picture, one can always write $2LH$ as a product of two triplet excitons. However, as was discussed earlier, the tight-binding picture cannot distinguish between singlet and triplet states energetically. Unsurprisingly, the electron-hole density correlation function for the tight-binding $2LH$ state is uniform, that is, up electrons are equally as likely to be correlated with down holes as they are with up holes. The degeneracy of the triplet and singlet states is lifted by the introduction of Coulomb repulsion via the Hubbard Hamiltonian, which also alters the density correlation for the $2LH$ state; the electrons and holes of the same (different) spin avoid (seek) each other. This feature only becomes apparent when observing the electron-hole density correlation functions rather than the electron density correlation function. 



While our calculations confirm that the $2LH$ state is indeed primarily composed of two electron-hole pairs and is close in energy to two times that of the triplet excitation, $2E(T_{1})$ \cite{zim}, it is not lower in energy than the $S_{1}$ state. It might be possible for this state to participate in the formation of two triplet excitons in the case of a pentacene dimer; however, in the monomer case it cannot be involved in MEG. Our calculation qualitatively agrees with that of another, more sophisticated, calculation \cite{dftcalc2}.
\section{\label{schi3} Two-Photon Absorption}

The linear optical properties of the -acenes are well known \cite{silb1,hexabs}. The linear absorption spectrum can be calculated by evaluating the imaginary component of the first-order optical susceptibility, $\chi^{(1)}(\omega)$ \cite{boyd}. Neglecting local field corrections and any solvent effects, if the molecules of interest are in solution, the contribution to the first-order susceptibility from solute molecules is given by
\beq
\chi_{\alpha\beta}^{(1)} \left(\omega \right)= \frac{N}{\epsilon_{0}\hbar} \sum_{n} \frac{\la \mu^{\alpha}_{gn} \mu^{\beta}_{ng}\ra}{\omega_{ng}-\omega - i\gamma_{ng} } + \frac{\la \mu^{\alpha}_{ng} \mu^{\beta}_{gn}\ra}{\omega_{ng}+\omega + i\gamma_{ng} }, \quad
\label{fosf}
\eeq
where $\alpha,\beta$ are Cartesian components, $N$ is the number density of molecules, $\epsilon_{0}$ is the vacuum permittivity, $\mu^{\alpha}_{nm}$ is the $\alpha^{th}$ component of the transition dipole moment between states $n$ and $m$, $\hbar\omega_{nm} = E_{n}-E_{m}$ is the energy difference between the states $n$ and $m$, and $\gamma_{nm}$ is the relaxation rate associated with these two states. By $\la \ra$ we indicate an average over molecular orientations; we consider a gas phase or solutions where the molecules are distributed randomly, so
\beq
\la \mu^{\alpha}_{gn}\mu^{\beta}_{ng} \ra =\frac{1}{3} \delta_{\alpha \beta} \sum_{\gamma}\mu^{\gamma}_{gn} \mu^{\gamma}_{ng},
\eeq
with the dipole moment matrix elements $\bso{\mu}_{gn}$ and $\bso{\mu}_{ng}$ lying in the plane of the molecule. The first-order susceptibility (\ref{fosf}) can be immediately computed from the information provided in Tables \ref{numtabpentethex} and \ref{oscstr}.
 
The nonlinear optical properties of the -acenes, such as two-photon absorption (TPA), have not been extensively explored in the literature; TPA allows for the investigation of dark states, such as the first doubly excited state. We compute the TPA of the -acenes by calculating the imaginary component of the third-order nonlinear optical susceptibility, $\chi^{(3)}(\omega; -\omega,\omega,\omega)$ \cite{boyd}. The largest contribution to the third-order nonlinear susceptibility is given by
\begin{widetext}
\beq
\chi_{\alpha\beta\gamma\delta}^{(3)}(\omega; -\omega,\omega,\omega) = \frac{N}{\epsilon_{0}\hbar^{3}} \mathcal{P}_{I} \sum_{vnm} \frac{\la \mu^{\alpha}_{gv}\mu^{\beta}_{vn}\mu^{\gamma}_{nm} \mu^{\delta}_{ml} \ra}{\left(\omega_{vg} - \omega -i\gamma_{vg} \right) \left(\omega_{ng} - 2\omega - i\gamma_{ng} \right) \left(\omega_{mg} - \omega - i\gamma_{mg} \right)},
\eeq
\begin{figure}[h]
\begin{center}
\includegraphics[scale=0.45]{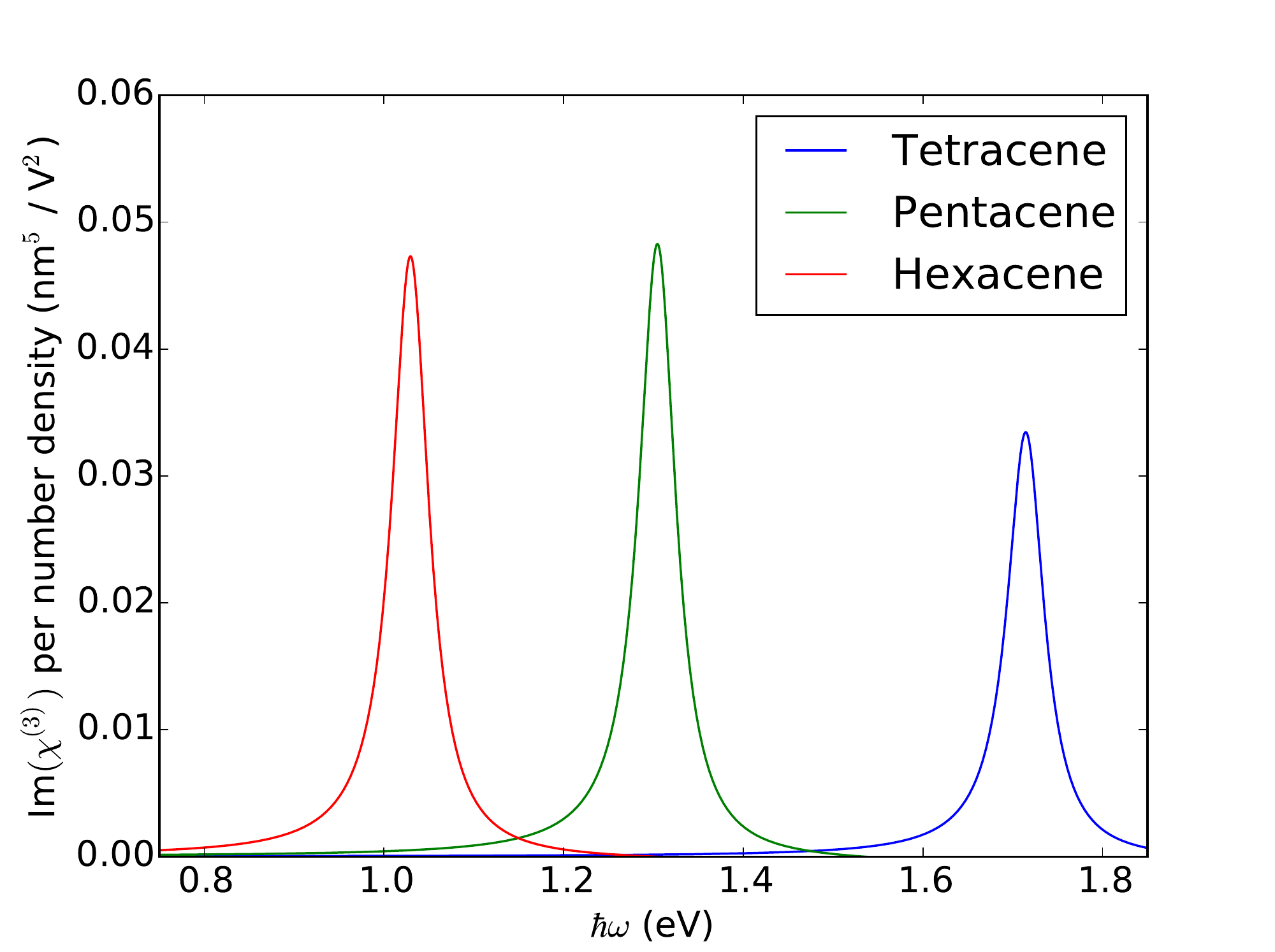}
\caption{Plot of $\text{Im} \left( \chi_{avg}^{(3)} \left( \omega;-\omega,\omega,\omega \right) \right)$ per number density for the three -acenes as a function of photon energy $\hbar \omega$; we have used units of nm$^{5}$ / V$^{2}$.}
\label{imchi3}
\end{center}
\end{figure}
\end{widetext}
where $\mathcal{P}_{I}$ is the permutation operator, and $\alpha,\beta,\gamma,\delta$ are Cartesian components. We set $\gamma_{nm} = \gamma = 0.05$ eV for all calculations; this is motivated by the linewidth of the $S_{1}$ absorption peak in the experimental absorption spectrum of pentacene \cite{silb1}. The indices $v$ and $m$ run over states that have a non-zero transition dipole moment with the ground state, these are the states $S_{k}$ where $k = 1,2,3,4$. The index $n$ runs over the states that have non-zero dipole moments with the $S_{k}$ states; in our model the only state that has a non-zero transition dipole moment with these singlets is the first doubly excited state, $2LH$. Again adopting a random distribution of molecular orientations we take
\begin{widetext}
\beq
&& \la \mu^{\alpha}_{gv}\mu^{\beta}_{vn}\mu^{\gamma}_{nm} \mu^{\delta}_{ml} \ra = \frac{1}{5} \delta_{\alpha\beta\gamma\delta}\sum_{\eta} \mu^{\eta}_{gv}\mu^{\eta}_{vn}\mu^{\eta}_{nm} \mu^{\eta}_{ml} \nonumber \\
&& + \frac{1}{15} \sum_{\eta \neq \eta'} \left(\delta_{\alpha\beta}\delta_{\gamma\delta}\mu^{\eta}_{gv}\mu^{\eta}_{vn}\mu^{\eta'}_{nm} \mu^{\eta'}_{ml} + \delta_{\alpha\gamma}\delta_{\beta\delta}\mu^{\eta}_{gv}\mu^{\eta'}_{vn}\mu^{\eta}_{nm} \mu^{\eta'}_{ml}+\delta_{\alpha\delta}\delta_{\beta\gamma}\mu^{\eta}_{gv}\mu^{\eta'}_{vn}\mu^{\eta'}_{nm} \mu^{\eta}_{ml}\right),
\label{spavg}
\eeq
\end{widetext}
where $\delta_{\alpha\beta\gamma\delta}$ is the generalized Kronecker delta function. We plot the full expression of $\text{Im}\left( \chi^{(3)}(\omega;-\omega,\omega,\omega) \right)$ per number density in Fig. \ref{imchi3}. The values are comparable to those calculated by other workers for the two-photon absorption of organic molecules such as coronene and hexa-peri-hexabenzocoronene (HBC) \cite{mazumdar}. The resulting two-photon absorption coefficients $\beta$ for tetracene, pentacene, and hexacene in solution are 25.10, 27.56, and 19.19 nm/GW respectively. where we assume the concentrations are those prepared in a previous experiment on pentacene \cite{penexpt}, and for the refractive index appearing in the expression for $\beta$ \cite{boyd} we have used the refractive index of the appropriate solvent.  
\section{\label{sconc} Conclusion}
%
%

In this paper we have demonstrated a computationally and physically simple scheme to extract the electronic excited state energies as well as wavefunctions of $\pi$ conjugated -acene molecules, specifically tetracene, pentacene and hexacene. A Hubbard model with a limited set of states was used to find these energies and wavefunctions. We have shown that the energies and oscillator strengths predicted by this model are in line with what one can achieve with modern quantum chemical techniques, but crucially without the computational complexity associated with these strategies. We also used our method to investigate the first doubly excited state, $2LH$, which is difficult to extract from quantum chemistry calculations. It has been speculated by some as being intimately involved in the singlet fission or multi-exciton generation process. 

We must emphasize that our simple calculations are not meant to replace more complex approaches; rather, the aim of our work is to generate a physically and computationally simple model that \textit{qualitatively} replicates the physics observed in the acenes molecules of tetracene, pentacene, and hexacene. Our goal is to use this simple technique to generate insight into the behavior of the electrons and holes in the excited states of these -acene molecules, especially those states that are hard to describe using traditional quantum chemistry techniques, such as the $2LH$ state.

We introduced a density correlation function, $g^{(2)}(s)$, to analyze the nature of these states. The ground state exhibits features typical of a noninteracting 1D electron gas, with oscillations in the $g^{(2)}(s)$ with a wavelength of $\lambda = 2l$, where $l$ is the bond length. The Hubbard interaction leads to the deepening of the correlation hole. The first doubly excited state, $2LH$, exhibits similar behavior. We found that the Hubbard interaction also deepens the correlation hole in triplets and singlets with respect to their tight-binding equivalents.

A more physically intuitive electron-hole picture was then introduced and we computed the density correlation of electrons with holes; these density correlation functions were then used to characterize the first doubly excited state, $2LH$. In the $2LH$ state, the electrons and holes of the same spin seek each other while the electrons and holes of different spins avoid each other. This type of behavior is reminiscent of triplet like excitons. While the $2LH$ state is seemingly composed of two triplet like excitons, it is higher in energy than the $S_{1}$ state and as such, at least in the case of a monomer, it cannot participate in multi-exciton generation. We have computed the two photon absorption of the -acenes and have shown there is strong two-photon absorption in the range of 1-1.8 eV in the -acenes; this absorption is due to excitation to the $2LH$ state.

\acknowledgments ZS thanks R. Schaffer, V. Venkataraman for edifying discussions.

\newpage
\begin{widetext}
\appendix

\section{\label{tothubbhamsupp}The Total Hamiltonian in the Electron-Hole Basis}

In the electron-hole basis the tight-binding Hamiltonian is
\begin{equation}
H_{TB} = \sum_{\substack{m\in\text{empty}, \\ \sigma}} \hbar\omega_{m} a^{\dagger}_{m\sigma} a_{m\sigma} - \sum_{\substack{m'\in\text{filled}, \\ \sigma}} \hbar\omega_{m'}b^{\dagger}_{m'\sigma} b_{m'\sigma} + 2\sum_{k \in\text{filled}} \hbar\omega_{k},
\end{equation}
while the Hubbard Hamiltonian can be written as
\begin{equation}
H_{Hu} =  \sum_{mm'pp'} \Gamma_{mm'pp'} C^{\dagger}_{m\uparrow}C_{m'\uparrow}C^{\dagger}_{p\downarrow}C_{p'\downarrow},
\end{equation}
where $\Gamma_{mm'pp'} = U\sum_{i} M^{*}_{im}M_{im'}M^{*}_{ip}M_{ip'}$. Moving to the electron-hole basis and normal ordering, the Hamiltonian can be expressed as
\begin{equation}
H = H_{0} +  H_{1} + H_{2} + H_{3} + H_{4}.
\end{equation}
The first component can be expressed as
\begin{equation}
H_{0} = 2\sum_{m\in\text{filled}} \hbar\omega_{m} + \sum_{\substack{l\in \text{filled}, \\ m \in \text{filled}}} \Gamma_{mmll} .
\label{H0} 
\end{equation}
The first part of Eq. \ref{H0} represents the tight-binding contribution of the nominal vacuum state $|0\rangle$ while the second part is the Coulomb repulsion of the tight-binding ground state.

The second part of the Hamiltonian is
\begin{eqnarray}
&& H_{1} = \sum_{m\sigma}\hbar\omega_{m}a^{\dagger}_{m\sigma}a_{m\sigma}  - \sum_{m'\sigma} \hbar\omega_{m'}b^{\dagger}_{m'\sigma}b_{m'\sigma} + \sum_{\substack{m \in \text{filled}, \\ pp' }} \Gamma_{mmpp'} (a^{\dagger}_{p\uparrow} a_{p'\uparrow} + a^{\dagger}_{p\downarrow} a_{p'\downarrow})  \label{H1} \\
&& - \sum_{\substack{m \in \text{filled}, \\ pp' }} \Gamma_{mmpp'} (b^{\dagger}_{p'\uparrow} b_{p\uparrow} +  b^{\dagger}_{p'\downarrow} b_{p\downarrow}). \nonumber
\end{eqnarray}
We have used the identity
\beq
\sum_{p \in \text{ filled}} \Gamma_{mm'pp} = 0,
\eeq
where $p$ ranges over the filled states, to simplify $H_{1}$. This identity is proved by
\begin{eqnarray}
&& \sum_{p \in \text{ filled}} \Gamma_{mm'pp} =  \sum_{i} \sum_{p  \in \text{ filled} } M^{*}_{ip}M_{ip} M^{*}_{im}M_{im'}, \\
&& = \frac{1}{2} \sum_{i} M^{*}_{im}M_{im'} = 0.
\end{eqnarray}
Eq. \ref{H1} represents the single particle terms that play a role in the matrix elements of both single and double excitations. The third part of Hamiltonian is
\begin{eqnarray}
&& H_{2} = \sum_{mm'pp'} \Gamma_{mm'pp'} \left( a^{\dagger}_{p\downarrow} b_{m\downarrow} - a^{\dagger}_{p\uparrow}b_{m\uparrow}\right)a_{m'\uparrow}a_{p'\downarrow} + \sum_{mm'pp'} \Gamma_{mm'pp'}a^{\dagger}_{m\uparrow}a^{\dagger}_{p\downarrow} \left( b^{\dagger}_{p'\uparrow}a_{m'\uparrow} - b^{\dagger}_{p'\downarrow}a_{m'\downarrow} \right), \label{H2} \\
&& \sum_{mm'pp'} \Gamma_{mm'pp'} \left( b^{\dagger}_{p'\downarrow} a_{m'\downarrow} - b^{\dagger}_{p'\uparrow}a_{m'\uparrow}\right)b_{p\uparrow}b_{m\downarrow} +  \sum_{mm'pp'} \Gamma_{mm'pp'}b^{\dagger}_{m'\downarrow}b^{\dagger}_{p'\uparrow} \left( a^{\dagger}_{m\downarrow}b_{p\downarrow} - a^{\dagger}_{m\uparrow}b_{p\uparrow} \right). \nonumber
\end{eqnarray}
While $H_{2}$ has matrix elements between tight-binding states that are in general non-zero, for the states and the active spaces investigated in this paper $H_{2}$ does not contribute. 
The fourth part of the Hamiltonian is
\begin{eqnarray}
&& H_{3} = -\sum_{mm'pp'} \Gamma_{mm'pp'}\left(a^{\dagger}_{p\downarrow}b^{\dagger}_{m'\downarrow}b_{m\downarrow}a_{p'\downarrow} +  a^{\dagger}_{p\uparrow}b^{\dagger}_{m'\uparrow}b_{m\uparrow}a_{p'\uparrow} \right),  \label{H3} \\ 
&& + \sum_{mm'pp'} \Gamma_{mm'pp'}\left( a^{\dagger}_{m\downarrow}b^{\dagger}_{m'\uparrow}b_{p\downarrow}a_{p'\uparrow} +  a^{\dagger}_{m\uparrow}b^{\dagger}_{m'\downarrow}b_{p\uparrow}a_{p'\downarrow} \right), \nonumber \\
&& - \sum_{mm'pp'} \Gamma_{mm'pp'}\left(b_{m\downarrow}b_{p\uparrow}a_{m'\uparrow}a_{p'\downarrow} +  a^{\dagger}_{m\uparrow}a^{\dagger}_{p\downarrow}b^{\dagger}_{m'\downarrow}b^{\dagger}_{p'\uparrow} \right). \nonumber
\end{eqnarray}
The term $H_{3}$ (Eq. (\ref{H3})) represents the part of the Hamiltonian that has a contribution to the matrix elements between single excitations, between the ground state and double excitations, as well as between different double excitations. 

The last part of the Hamiltonian is written as
\begin{eqnarray}
&& H_{4} = \sum_{mm'pp'}\Gamma_{mm'pp'} a^{\dagger}_{p\downarrow}a^{\dagger}_{m\uparrow}a_{m'\uparrow}a_{p'\downarrow} + \sum_{mm'pp'}\Gamma_{mm'pp'} b^{\dagger}_{m'\downarrow}b^{\dagger}_{p'\uparrow}b_{p\uparrow}b_{m\downarrow}.
\label{H4}
\end{eqnarray}
$H_{4}$ (Eq. (\ref{H4})) is the part of the Hamiltonian which contributes to the matrix elements between double excitations only.

\section{\label{dipsupp} Dipole Operator in the electron-hole Basis}
Neglecting overlap between neighboring $p_{z}$ orbitals, the dipole moment operator of a molecule with nuclei assumed fixed is given by
\begin{equation}
\bso{\mu}= \sum_{i} e \mbf{r_{i}} \left(\sum_{\sigma} c^{\dg}_{i\sigma} c_{i\sigma} - 1 \right),
\end{equation}
where $\mbf{r_{i}}$ is the position of site $i$ from any chosen origin, $e = -|e|$ is the electronic charge, and the charge of each nucleus not balanced by the in-plane bonding electrons of the molecule is included so the dipole moment operator is independent of origin. Transforming to the tight-binding basis, the $\alpha^{th}$ Cartesian component of the dipole moment is
\begin{equation}
\mu^{\alpha} = \left( \sum_{mm'}{\mu^{\alpha}_{mm'}(C^{\dagger}_{m\uparrow}C_{m'\uparrow}+C^{\dagger}_{m\downarrow}C_{m'\downarrow})} \right) - \sum_{i}er^{\alpha}_{i},
\label{dipopTB}
\end{equation}
where the operators $C^{\dg}_{m\sigma}$ and $C_{m\sigma}$ are defined by Eqs. (\ref{TBCdg}, \ref{TBC}) respectively. The quantity $\mu^{\alpha}_{mm'}$ is the $\alpha^{th}$ component of the electronic contribution to the dipole matrix element between two tight binding states $m$ and $m'$ which is
\beq
\mu^{\alpha}_{mm'} = e\sum_{i} r^{\alpha}_{i} M^{*}_{im} M_{im'},
\eeq
where the sum ranges over all sites $i$, $r^{\alpha}_{i}$ is the position of the $\alpha^{th}$ coordinate of site $i$, and $M_{im}$ is the amplitude of the $i^{th}$ site for the $m^{th}$ tight-binding state. In the electron-hole picture, the $\alpha^{th}$ component of the dipole operator (\ref{dipopTB}) is
\begin{eqnarray}
&& \mu^{\alpha} + \sum_{i} er^{\alpha}_{i}= \mu^{\alpha}_{1} + \mu^{\alpha}_{2} + \mu^{\alpha}_{3},
\end{eqnarray}
where
\begin{eqnarray}
&& \mu^{\alpha}_{1} = \sum_{mm'}{\mu^{\alpha}_{mm'}(a^{\dagger}_{m\uparrow}a_{m'\uparrow} + a^{\dagger}_{m\downarrow}a_{m'\downarrow})}, \\
&& \mu^{\alpha}_{2} =  - \sum_{mm'}{\mu^{\alpha}_{mm'}(b^{\dagger}_{m'\uparrow}b_{m\uparrow} + b^{\dagger}_{m'\downarrow}b_{m\downarrow})}, \\
&& \mu^{\alpha}_{3} = \sum_{mm'}{\mu^{\alpha}_{mm'}(a^{\dagger}_{m\uparrow}b^{\dagger}_{m'\downarrow} + a^{\dagger}_{m\downarrow}b^{\dagger}_{m'\uparrow} + b_{m'\downarrow}a_{m\uparrow} + b_{m'\uparrow}a_{m\downarrow})}.
\end{eqnarray}

\newpage
\section{\label{activespace} Changing the number of basis states and its impact on transition energies and density correlation functions}
We have investigated the effect of both increasing and decreasing the size of the basis on the transition energies and density correlation functions of pentacene. The basis we use in the text we denote as the ``full active space" or FAS. This is composed of all single excitations of the form of Eq. (\ref{singbasis}) and all double excitations of the form of Eq. (\ref{dubbasis}); the levels involved are $H_{3}$ to $L_{3}$. 

\begin{figure}[h]
\begin{center}
\includegraphics[scale=0.44]{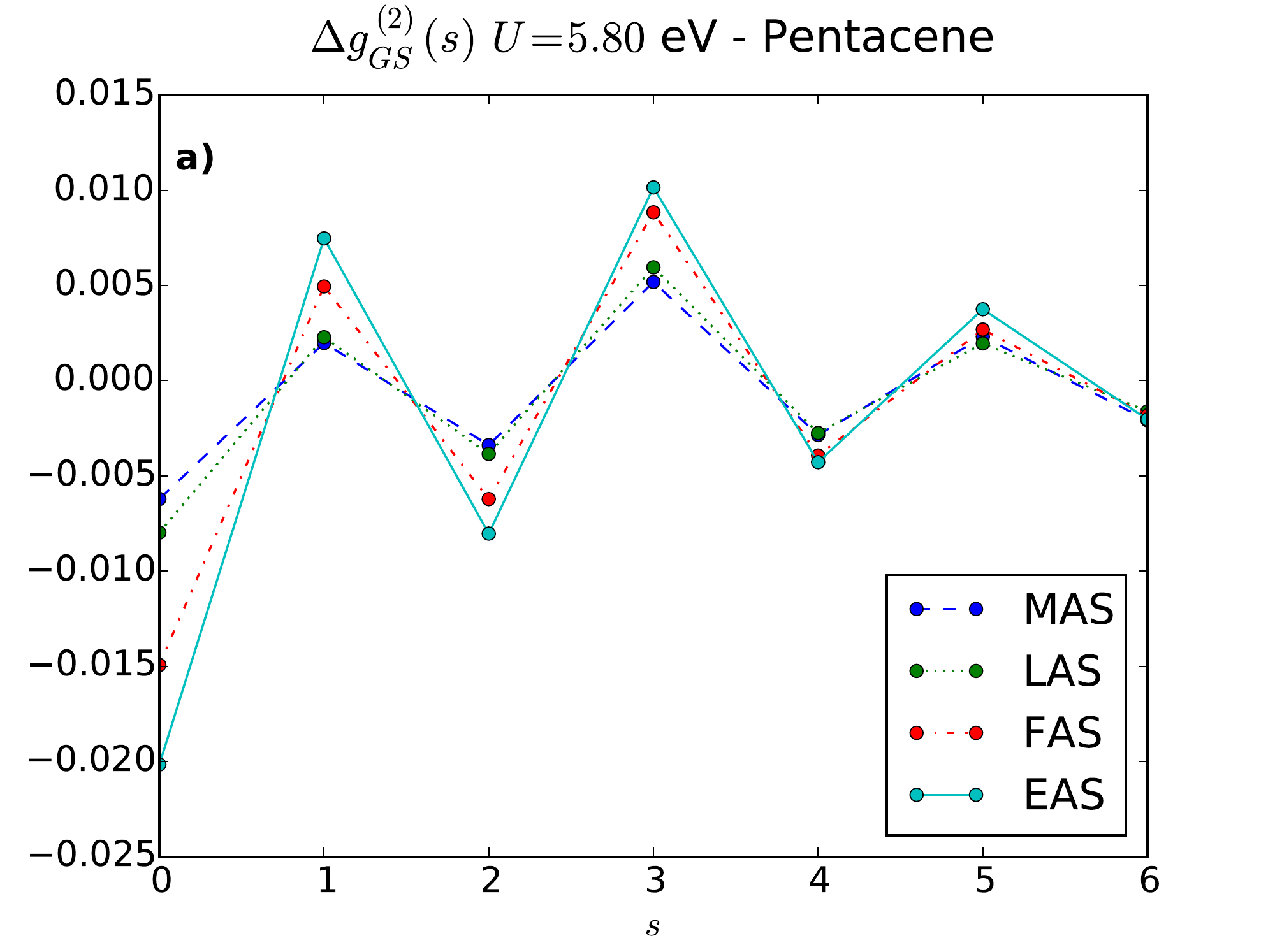}
\includegraphics[scale=0.44]{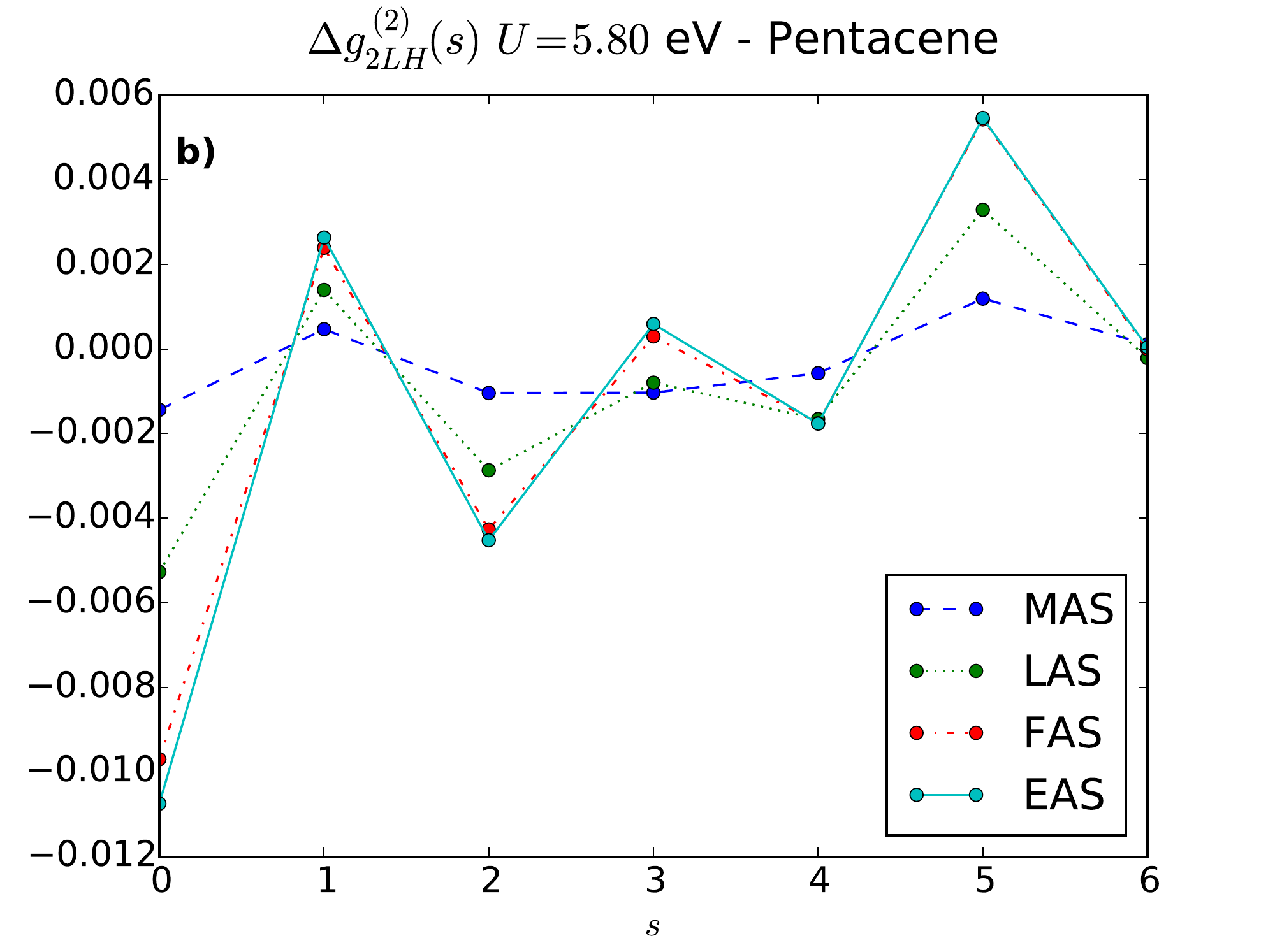}
\caption{Plots of a) $\Delta g_{GS}^{(2)}(s)$ and b) $\Delta g_{2LH}^{(2)}(s)$ of pentacene for four different active spaces, MAS, LAS, FAS, and EAS. Qualitatively, decreasing the size of the basis has little impact on the density correlation of the ground and the $2LH$ state. In these plots, we fix the value of the Hubbard parameter at $U = 5.80$ eV, the value for which the first singlet transition matches the literature value for the FAS basis.}
\label{corrasfig}
\end{center}
\end{figure}

\newpage
\begin{figure}[h]
\begin{center}
\includegraphics[scale=0.45]{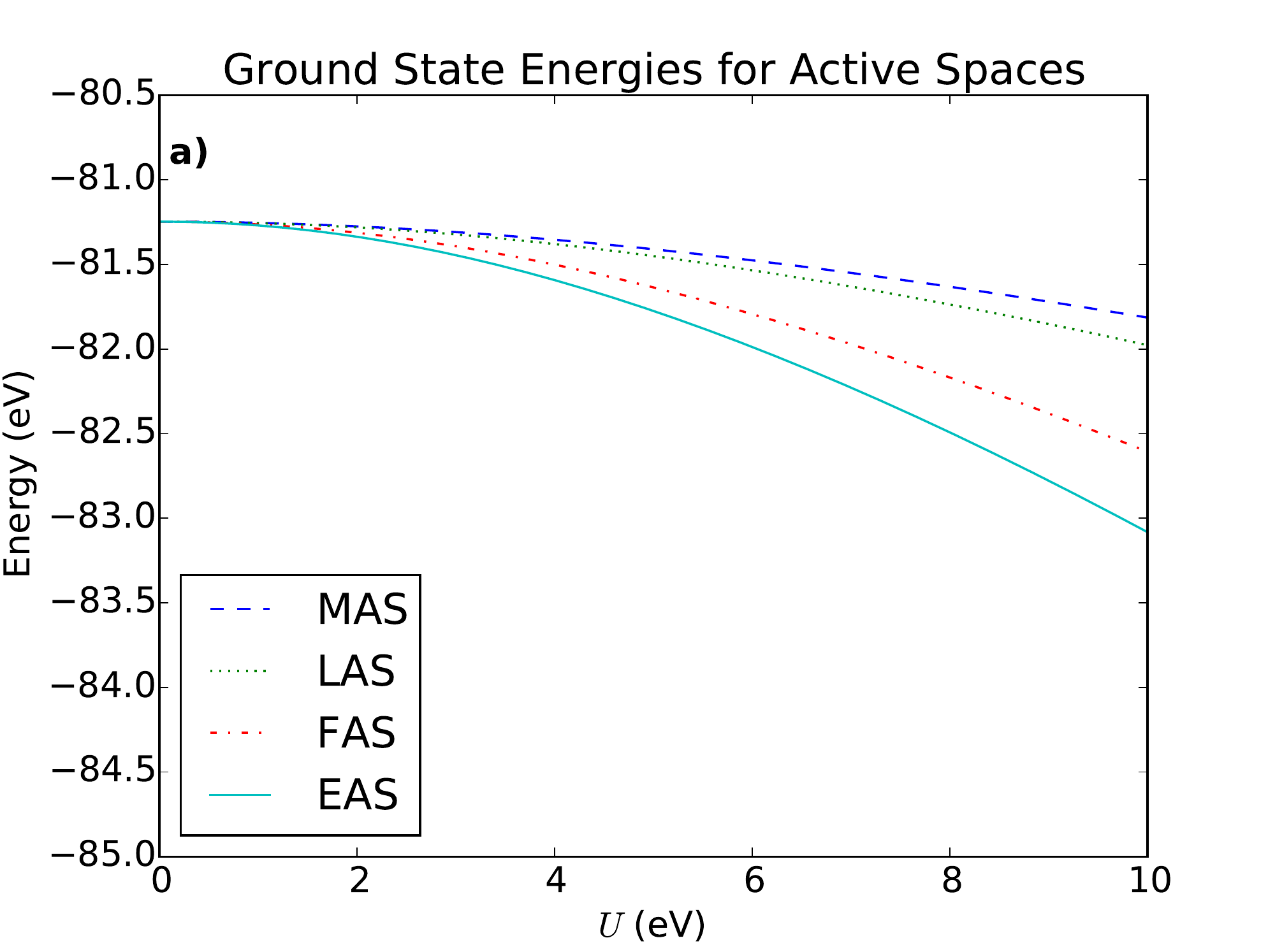}
\includegraphics[scale=0.45]{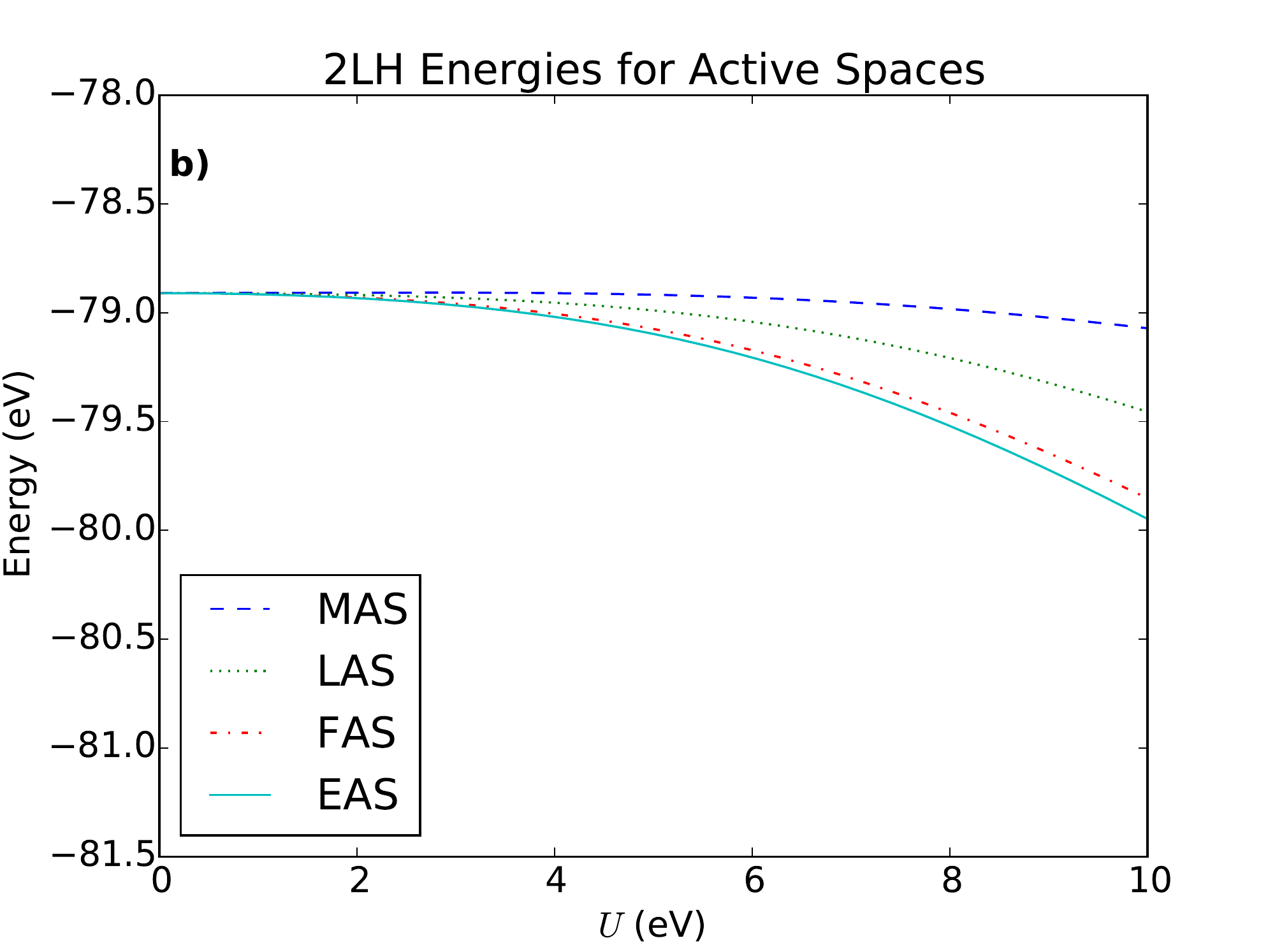}
\caption{a) Absolute energies of the ground and b) $2LH$ state plotted as a function of $U$ for various active spaces.}
\label{asae}
\end{center}
\end{figure}

We have considered two smaller basis sets and one larger one. The first of these smaller bases consists of all single excitations of the form of Eq. (\ref{singbasis}) and all the double excitations of the form Eq. (\ref{dubbasis}), but involving only the levels $H_{1}$ to $L_{1}$. We call this basis set the ``limited active space", or LAS. The other smaller basis set we used consists of select single and double excitations from the LAS. This basis is called the ``minimum active space", or MAS. The double excitations included in the MAS are
\begin{eqnarray}
&& |\overline{2LH}\rangle = a^{\dagger}_{L\uparrow}a^{\dagger}_{L\downarrow}b^{\dagger}_{H\downarrow}b^{\dagger}_{H\uparrow}|0\rangle, \label{pasbasis1} \\
&& |\overline{2LH_{1}} \rangle = a^{\dagger}_{L\uparrow}a^{\dagger}_{L\downarrow}b^{\dagger}_{H_{1}\downarrow}b^{\dagger}_{H_{1}\uparrow}|0\rangle, \\
&& |\overline{2L_{1}H} \rangle = a^{\dagger}_{L_{1}\uparrow}a^{\dagger}_{L_{1}\downarrow}b^{\dagger}_{H\downarrow}b^{\dagger}_{H\uparrow}|0\rangle, \\
&& |\overline{2L_{1}H_{1}} \rangle = a^{\dagger}_{L_{1}\uparrow}a^{\dagger}_{L_{1}\downarrow}b^{\dagger}_{H_{1}\downarrow}b^{\dagger}_{H_{1}\uparrow}|0\rangle, \\
&& |\overline{LL_{1};HH_{1}} \rangle =  a^{\dagger}_{L\uparrow}a^{\dagger}_{L_{1}\downarrow}b^{\dagger}_{H\downarrow}b^{\dagger}_{H_{1}\uparrow}|0\rangle,\\
&& |\overline{L_{1}L;H_{1}H} \rangle =  a^{\dagger}_{L_{1}\uparrow}a^{\dagger}_{L\downarrow}b^{\dagger}_{H_{1}\downarrow}b^{\dagger}_{H\uparrow}|0\rangle,
\label{pasbasis1a}
\end{eqnarray}
and single excitations included are
\begin{eqnarray}
&& |\overline{LH,\sigma}\rangle = a^{\dagger}_{L\sigma}b^{\dagger}_{H\tilde{\sigma}}|0\rangle, \label{pasbasis2a} \\
&& |\overline{L_{1}H_{1},\sigma}\rangle = a^{\dagger}_{L_{1}\sigma}b^{\dagger}_{H_{1}\tilde{\sigma}}|0\rangle, \\
&& |\overline{LH_{1}, \sigma}\rangle =  a^{\dagger}_{L\sigma}b^{\dagger}_{H_{1}\tilde{\sigma}}|0\rangle, \\
&& |\overline{L_{1}H, \sigma}\rangle =  a^{\dagger}_{L_{1}\sigma}b^{\dagger}_{H\tilde{\sigma}}|0\rangle, \\
&& |\overline{L_{2}H, \sigma}\rangle =  a^{\dagger}_{L_{2}\sigma}b^{\dagger}_{H\tilde{\sigma}}|0\rangle, \\
&& |\overline{L_{3}H, \sigma}\rangle =  a^{\dagger}_{L_{3}\sigma}b^{\dagger}_{H\tilde{\sigma}}|0\rangle, \\
&& |\overline{LH_{2}, \sigma}\rangle =  a^{\dagger}_{L\sigma}b^{\dagger}_{H_{2}\tilde{\sigma}}|0\rangle, \\
&& |\overline{LH_{3}, \sigma}\rangle =  a^{\dagger}_{L\sigma}b^{\dagger}_{H_{3}\tilde{\sigma}}|0\rangle.
\label{pasbasis2}
\end{eqnarray}
The larger basis consists of all single excitations of the form of Eq. (\ref{singbasis}) and all the double excitations of the form Eq. (\ref{dubbasis}), involving only the levels $H_{4}$ to $L_{4}$. We call this basis the ``extended active space", or EAS.
%
%

The density correlation functions for the ground and $2LH$ state are plotted in Fig. \ref{corrasfig} for a value of $U$ such that the $S_{1}$ transition energy matches the literature value in pentacene for the FAS basis. As we reduce the size of the basis, there is little qualitative difference in the density correlation function of these states. As we increase the size of the basis, we can see that there is very little difference in the density correlation functions between the active space used in the text and those of the extended active space, suggesting that the ground state and $2LH$ state have essentially converged in the FAS basis. 

The transition energies for the various basis sets are shown in Table \ref{acspte}. The absolute energies of the ground and $2LH$ state are plotted in Fig. \ref{asae}. 

\begin{center}
\begin{table}[h]
\begin{tabular}{|c|c|c|c|c|c|}
\hline 
\textbf{Basis} & $U$ \textbf{(eV)} & $T_{1}$ \textbf{(eV)} & $2LH$ \textbf{(eV)} & $g\rightarrow S_{1}$ \textbf{oscillator strength} & $S_{1}\rightarrow2LH$ \textbf{oscillator strength}\tabularnewline
\hline 
\hline 
MAS & 7.53 & 0.74 & 2.63 & 0.127$(\hat{\mathbf{y}})$ & 0.102$(\hat{\mathbf{y}})$\tabularnewline
\hline 
LAS & 7.00 & 0.84 & 2.52 & 0.131$(\hat{\mathbf{y}})$ & 0.0691$(\hat{\mathbf{y}})$\tabularnewline
\hline 
FAS & 5.80 & 1.06 & 2.61 & 0.146$(\hat{\mathbf{y}})$ & 0.0877$(\hat{\mathbf{y}})$\tabularnewline
\hline 
EAS & 5.15 & 1.17 & 2.69 & 0.178$(\hat{\mathbf{y}})$ & 0.111$(\hat{\mathbf{y}})$\tabularnewline
\hline 
\end{tabular}
\caption{Transition energies of the first triplet and the first doubly excited state, along with oscillator strengths for the $g \rightarrow S_{1}$ and $S_{1} \rightarrow 2LH$ transition,  of pentacene for various basis sets. The direction of the transition dipole moment of these transitions are indicated in parenthesis; the molecular axes are shown in Fig. \ref{pendir}. We choose a $U$ such that the first singlet transition energy matches the experiment; hence for each basis the value of $U$ is different.}
\label{acspte}
\end{table}
\end{center}

\newpage
\section{\label{physmean} Electron Density Correlations of $2LH$, $S_{1}$, and $T_{1}$ states Upon the Introduction of Coulomb Repulsion}

\begin{figure}[h]
\begin{center}
\includegraphics[scale=0.45]{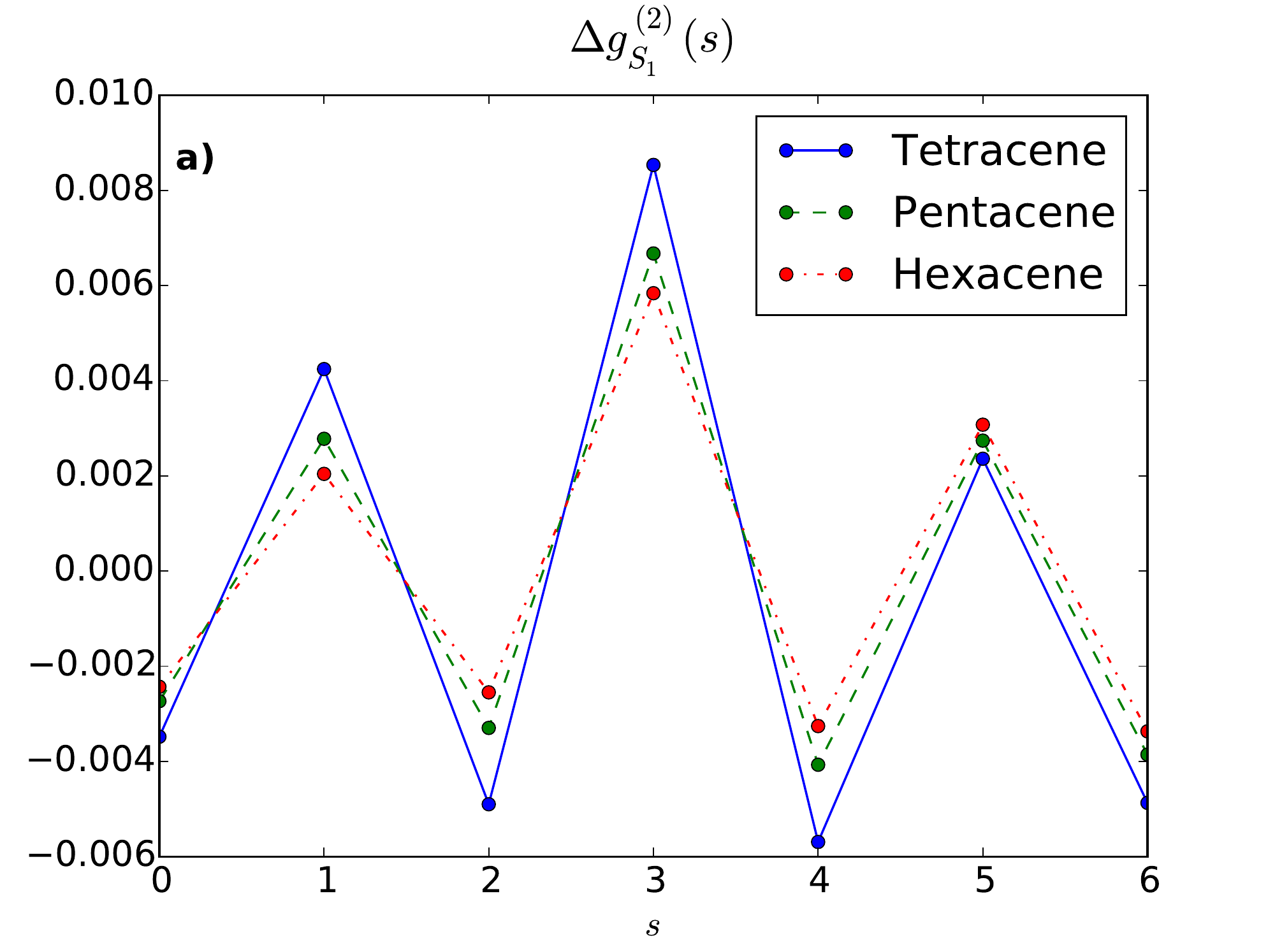}
\includegraphics[scale=0.45]{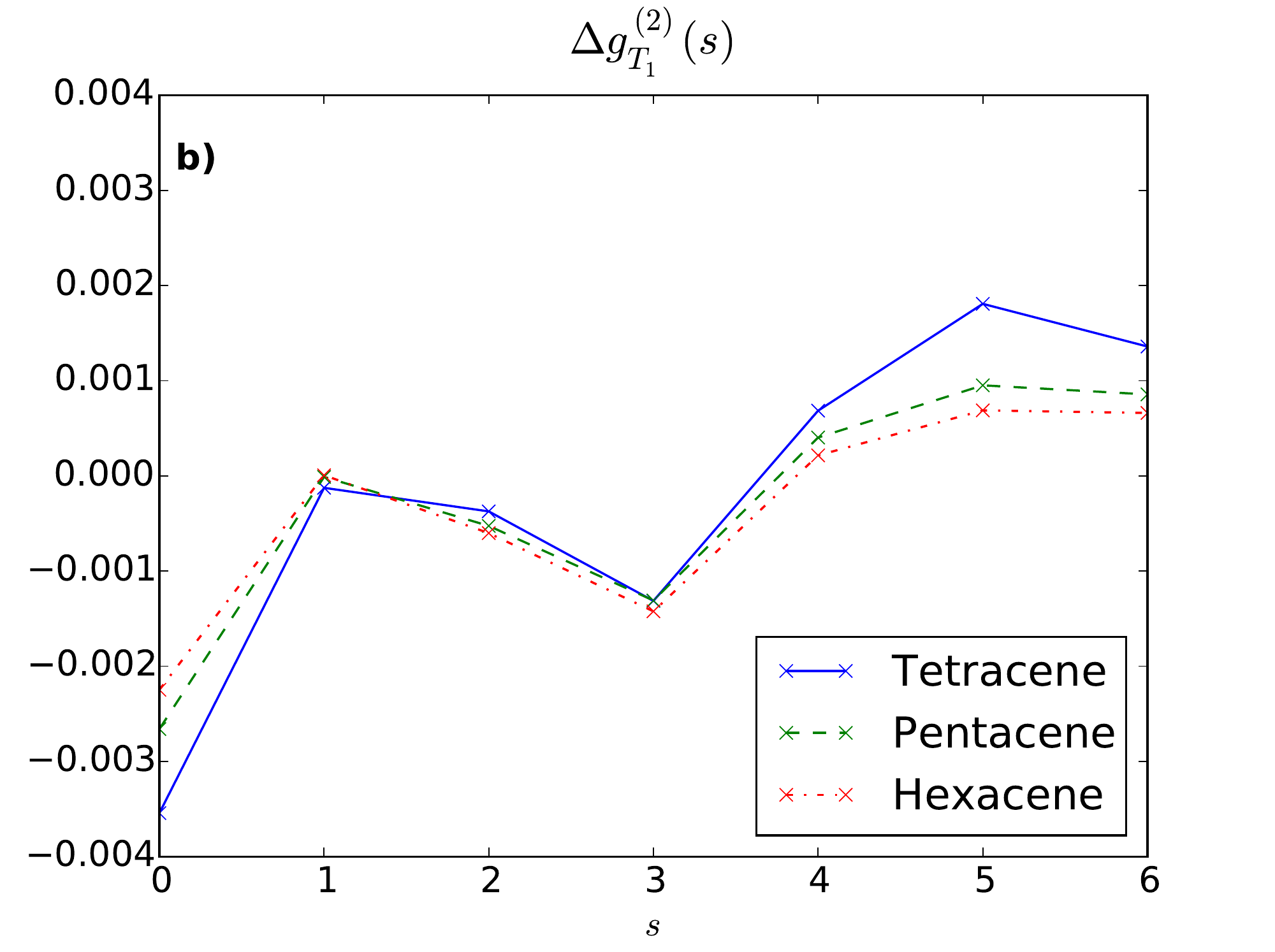}
\caption{Plots of a) $\Delta g_{S_{1}}^{(2)}(s)$, and b) $\Delta g_{T_{1}}^{(2)}(s)$ in tetracene, pentacene, and hexacene.  Upon the introduction of the Coulomb repulsion, the Fermi hole deepens for both the singlet and triplet state relative to their respective tight-binding states.} 
\label{delg1comp}
\end{center}
\end{figure}

\begin{figure}[h]
\begin{center}
\includegraphics[scale=0.48]{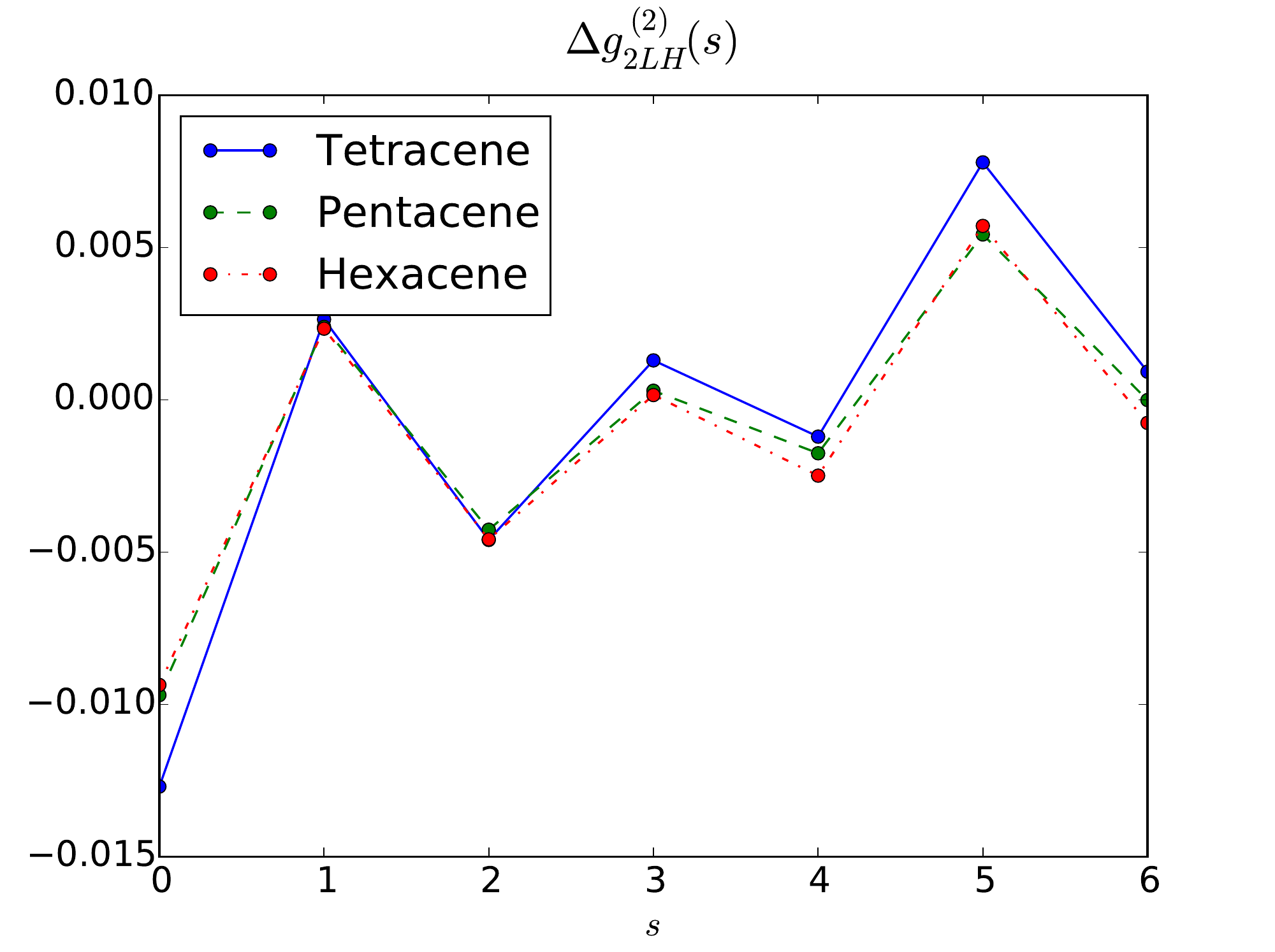}
\caption{$\Delta g_{2LH}^{(2)}(s)$ plotted for tetracene, pentacene, and hexacene. The Hubbard interaction in this case stabilizes the $2LH$ state by deepening the Fermi hole.}
\label{delg12HL}
\end{center}
\end{figure}

The quantity $\Delta g_{X}^{(2)}(s)$ (\ref{Deltag1}) is plotted for $X= S_{1}$, and $T_{1}$ for all three -acenes in Fig. \ref{delg1comp}. Upon the introduction of the Coulomb repulsion, the Fermi hole deepens for both the triplet and singlet states. The $\Delta g^{(2)}_{2LH}(s)$ quantity is also plotted for all three -acenes in Fig. \ref{delg12HL}; the introduction of the Coulomb repulsion also leads to a deepening of the Fermi hole.  

\newpage
\end{widetext}

\end{document}